\documentclass[pra,showpacs,psfig]{revtex4}

\usepackage{amssymb}
\usepackage[dvips]{graphicx} 
\usepackage{pstricks} 
\newtheorem{3by3}{Theorem}
\newtheorem{GT1}[3by3]{Theorem}
\newtheorem{GT2}[3by3]{Theorem}
\newtheorem{NCerf}[3by3]{Theorem}
\newtheorem{SEP}[3by3]{Theorem}
\begin{document}

\title{Understanding entanglement as resource: locally distinguishing unextendible product bases}
\author{Scott M. Cohen$^{1,2}$}
\email{cohensm@duq.edu}
\affiliation{$^1$Department of Physics, Duquesne University,
Pittsburgh, Pennsylvania 15282\\
$^2$Department of Physics, Carnegie-Mellon University,
Pittsburgh, Pennsylvania 15213}

\begin{abstract}
It is known that the states in an unextendible product basis (UPB) cannot be distinguished perfectly when the parties are restricted to local operations and classical communication (LOCC). Previous discussions of such bases have left open the following question: What entanglement resources are necessary and/or sufficient for this task to be possible with LOCC? In this paper, I present protocols which use entanglement more efficiently than teleportation to distinguish certain classes of UPB's. The ideas underlying my approach to this problem offer rather general insight into why entanglement is useful for such tasks.
\end{abstract}

\maketitle

\section{Introduction}
One of the most fascinating aspects of quantum physics is the possibility that spatially separated systems may be ``entangled" with each other, exhibiting correlations that have no counterpart, or explanation, in classical physics. This characteristic of quantum states has long stimulated lively debate amongst physicists and others about the implications of such correlations for our understanding of the world around us. Beyond such philosophical questions, however, entanglement has in recent years been shown to be a valuable resource, allowing remote parties to communicate in ways that were previously not thought possible. Examples include the well-known protocols of teleportation \cite{BennettTele} and dense coding \cite{BennettDense}, which have helped spawn rapid growth in the relatively new field of quantum information. The discovery of the potential power of quantum computers \cite{Shor_expand}, which may also rely on entanglement, has been an additional motivating factor stimulating this period of growth.

It is of fundamental interest to understand the extent to which entanglement can expand the reach of what it is possible to accomplish. The most important questions in this regard concern spatially separated parties performing quantum operations on their local systems, and possibly communicating classical information to each other. We are thus led to consider a restriction to local operations and classical communication, commonly denoted as LOCC.

One widely studied question is the following: Suppose a collection of quantum systems in some chosen state is distributed to the parties. The parties are not told the state their combined system is in, but they know it has been chosen from a specific set of mutually orthogonal states, that set being known to each of them. Their task is to determine the chosen state by using LOCC, a task we may refer to as local state discrimination. Given a particular such set of states, we may wish to know if it is possible for the parties to perfectly distinguish the state; that is, given any one of the states in the set and by using LOCC, can they with certainty determine which state they were given? In the remainder of this paper, when we refer to such distinguishability questions, they should be understood to imply this notion of perfect distinguishability.

A fascinating result related to these questions was the discovery by Bennett, et. al. \cite{Bennett9} of a complete basis of product (i.e., unentangled) states that cannot be distinguished by LOCC. The system under consideration involved two parties, with both parties' systems having dimensionality $3$ (more concisely, $3\otimes3$). The authors dubbed this phenomenon ``nonlocality without entanglement" (NLWE), and various simplifications of their proof have since appeared in the literature \cite{GroismanVaidman,WalgateHardy,myLDPE}. Their construction of complete bases was later generalized \cite{NisetCerf} to the case of many parties with arbitrary dimensions of their Hilbert spaces.

There also exist what might be termed ``incomplete bases" that exhibit this effect. Suppose one has a set of mutually orthogonal product states satisfying the condition that no product state lies in the orthogonal complement of the subspace spanned by these states. Then this set cannot be extended, in the sense that no additional product state can be added to it while preserving the orthogonality of the set. Such a set of product states is known as an ``unextendible product basis", or UPB \cite{IBM_PRL,IBM_CMP,NisetCerf,AlonLovasz,Rinaldis}. These sets are of considerable interest within the quantum information community, not only because they exhibit NLWE, but also because the (normalized) projector onto that orthogonal complement is a mixed state that exhibits the fascinating phenomenon known as bound entanglement \cite{HorodeckiBound,HorodeckisBound}.

Any set of states that cannot be distinguished by LOCC alone can nonetheless always be distinguished by LOCC if the parties share enough entanglement. That is, with enough entanglement, LOCC can be used to teleport \cite{BennettTele} the full multipartite state to a single party, and that party can then make a measurement to determine which state they were given. However, it is by now widely understood that entanglement is a valuable resource, so it is important to ask if this task can be accomplished more efficiently. This is the question we address in the present paper.

Specifically, we will consider several examples of UPB's and devise LOCC protocols that, using entanglement as a resource, distinguish these sets considerably more efficiently than teleportation. Each of the UPB's we consider is closely related to a complete NLWE basis, and the latter can also be distinguished by the same (or slightly altered) protocols. Perhaps just as important is the method we use to devise these protocols, which is described in detail below. The approach involves the use of box diagrams to represent the set of states to be distinguished. Such diagrams have been used in a related context before \cite{Bennett9}, but as far as we are aware, they have not been used to follow the dynamics of the systems being represented, which is what we have done here. We have found these diagrams to be extremely useful as a represention of the set of states \textit{in conjunction with entanglement}. In particular, we take the perspective that entanglement provides the parties with \textit{multiple images} of the set of states to be distinguished, and it is this perspective that has led us to our protocols. We believe this perspective will prove useful in other applications, as well.

The paper is organized as follows: In the following section, we introduce the box diagrams and explain how they may be used to represent quantum states and to follow the effects of measurements and other operations. We also explain what is meant by the idea of ``multiple images". Section \ref{sec:3by3} should serve as an introduction to the general ideas with regard to how the protocols are constructed. In this section, UPB's on $3\otimes3$ systems are considered, including a proof that all such UPB's can be disinguished by LOCC with a $2\otimes2$ maximally entangled state (MES). The following section considers a UPB on a higher dimensional Hilbert space, specifically the GenTiles1 UPB \cite{IBM_CMP,DiVincenzoTerhal} on a $6\otimes6$ system. In an appendix, it is shown that every UPB in the GenTiles1 family is distinguishable with an $m/2\otimes m/2$ MES, where $m$ is the Hilbert space dimension for each party. Similarly, in another appendix, the GenTiles2 family of UPB's \cite{IBM_CMP,DiVincenzoTerhal} on $m\otimes n$ with $n\ge m$, is shown to be distinguishable with a $\lceil m/2\rceil \otimes \lceil m/2\rceil$ MES, where $\lceil x\rceil$ is defined to be the smallest integer not less than $x$. In Section \ref{sec:NisetCerf}, we show that every UPB in the Niset $\&$ Cerf construction for many parties \cite{NisetCerf} is LOCC-distinguishable with a single $2\otimes2$ MES shared between any two of them, regardless of how many parties or the dimensions of their Hilbert spaces. Finally, in Section~\ref{sec:disc}, we discuss these results and then end with conclusions.

\section{Visualizing quantum information processing}\label{sec:vqip}
In subsequent sections, we will give detailed protocols which locally distinguish specific unextendible product bases. In this section, we will motivate these protocols with simple box diagrams, which have been used for other studies in quantum information processing. As far as we are aware, however, our use of these diagrams to follow the dynamics of the systems involved is new. One of the crucial observations will be that a shared entangled state (on systems $a$ and $b$, for example) provides the parties with multiple ``images" of the state of an additional system, say $B$. As we will see, the way these images are distributed through Hilbert space is such that the parties, independently of each other, are able to manipulate the individual images, and they can manipulate each image differently from the other images. This idea is the basis underlying our protocols for distinguishing UPB's. We have chosen to use the term ``images" rather than copies, because the latter would tend to imply the notion of clones, which of course are impossible to create \cite{NoCloning,cloneDieks}. See below for further clarification on these points. The following discussion begins at an elementary level to provide the reader with ample opportunity to understand the general ideas. These ideas have been presented elsewhere at an even more elementary level, accessible to undergraduates and perhaps others, in the context of ``visualizing teleportation" \cite{my_tele}.

\subsection{Quantum states}
We begin with simple illustrations of how box diagrams can be used to represent quantum states. As an example, an arbitrary state $\vert\Psi\rangle_B$ on a two-level (qubit) system $B$ is

\begin{equation}
\includegraphics{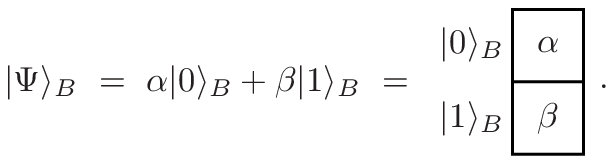}
\end{equation}

If a second qubit $a$ is in the state $\vert 0\rangle_a$, the diagram becomes
\begin{equation}
\includegraphics{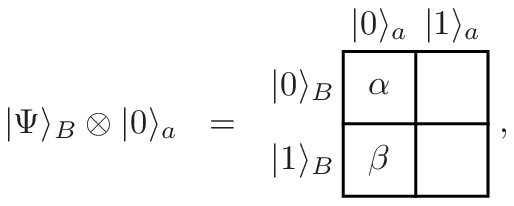}
\end{equation}
where the empty squares on the right-hand side of this diagram represent the fact that system \textit{a} is in state $\vert 0\rangle_a$ and not in $\vert 1\rangle_a$.

If there are three systems involved, a three-dimensional cube could be used to represent this situation. However, it will serve our purposes to represent two of these systems, when they are both held by Bob, along the vertical dimension of the diagram. If his second system $b$ is also in the $\vert 0\rangle_b$ state, then
\begin{equation}
\includegraphics{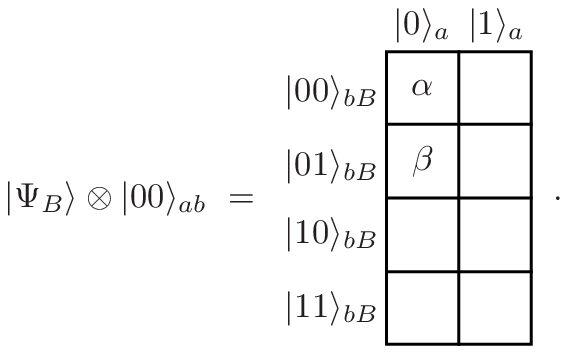}
\end{equation}
If instead the \textit{a,b} systems are both $\vert 1\rangle$, this picture is
\begin{equation}
\includegraphics{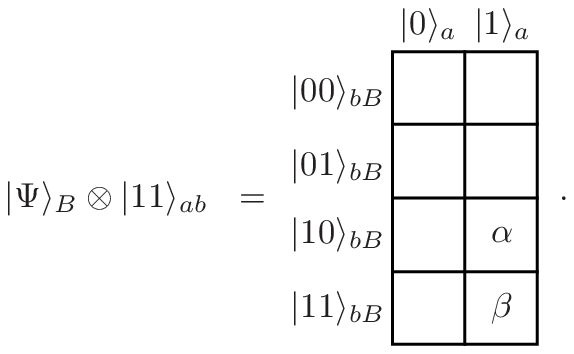}
\end{equation}
If systems $a,b$ are in the ``maximally entangled state" state $|B_0\rangle_{ab} = |00\rangle_{ab} + |11\rangle_{ab}$ the corresponding diagram is simply the sum of the previous two; that is,
\begin{equation}\label{B0_dgrms}\label{B0SA}
\includegraphics{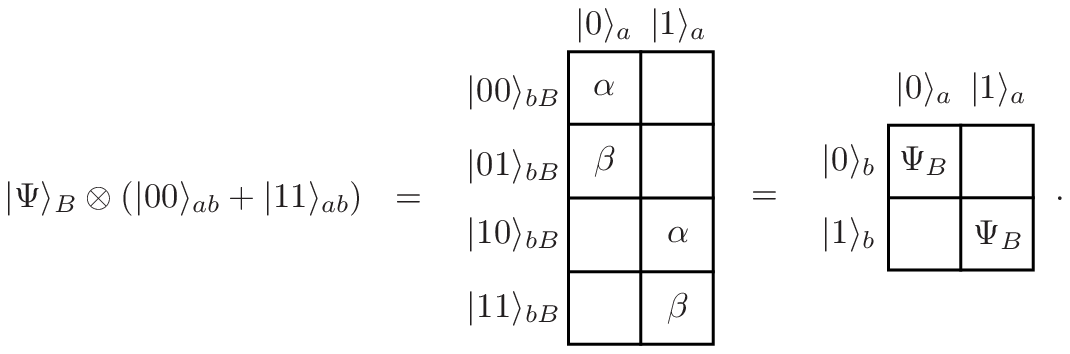}
\end{equation}
Notice how there are now two images of the state $|\Psi\rangle_B$. Clearly, if the entangled state on $a,b$ was of higher Schmidt rank, say $r$, then there would be $r$ images of $|\Psi\rangle_B$. We note once again that these are not ``clones", since the ``images" of the state $|\Psi\rangle$ are all on the same system $B$.

Let us next look at how to represent measurements by use of these diagrams.

\subsection{Measurements on quantum systems}
Suppose Bob and Alice share the three quantum systems in the state represented in Eq.~(\ref{B0SA}). Bob could make a measurement in the standard basis on system $b$. If he obtains outcome $\vert 0\rangle_b$, this would be represented as
\begin{equation}
\includegraphics{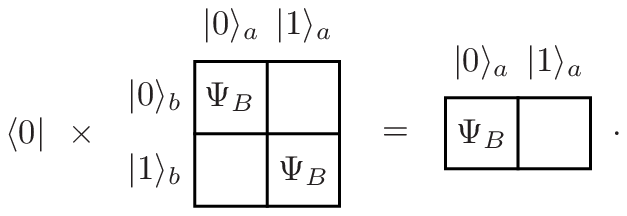}
\end{equation}
If instead his outcome had been $\vert 1\rangle_b$, we would have
\begin{equation}
\includegraphics{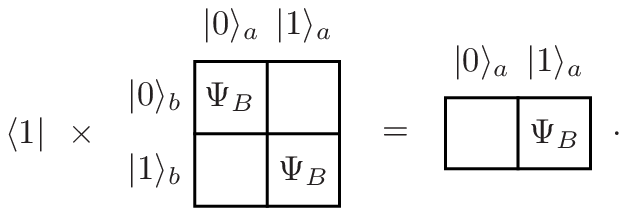}
\end{equation}
Alternatively, he could do a measurement in the $\vert\pm\rangle_b = \vert 0\rangle_b \pm \vert 1\rangle_b$ basis, and if his outcome is $\vert +\rangle_b$, we have
\begin{equation}
\includegraphics{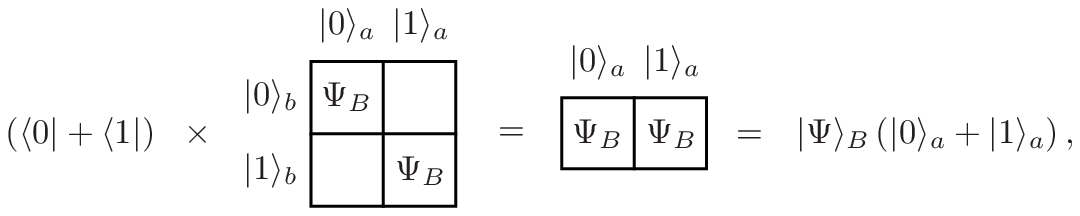}
\end{equation}
which is just a sum of the previous two equations (notice how the two images have collapsed into a single row as a result of this rank-1 measurement outcome). We see that in each of these cases, the state of system $a$ is directly correlated to the outcome of Bob's measurement on system $b$, which is an obvious consequence of the fact these two systems started out in the entangled state $\vert B_0\rangle_{ab}$.

The way the images of $\Psi_B$ appear in the diagram is crucial: that the two start out in different rows \textit{and} in different columns will be important in what is to come. If entanglement between systems \textit{a,b} was absent, for example if they were in the (unentangled) state $|+\rangle_b\otimes|0\rangle_a$, then this would be represented by
\begin{equation}\label{obscure}
\includegraphics{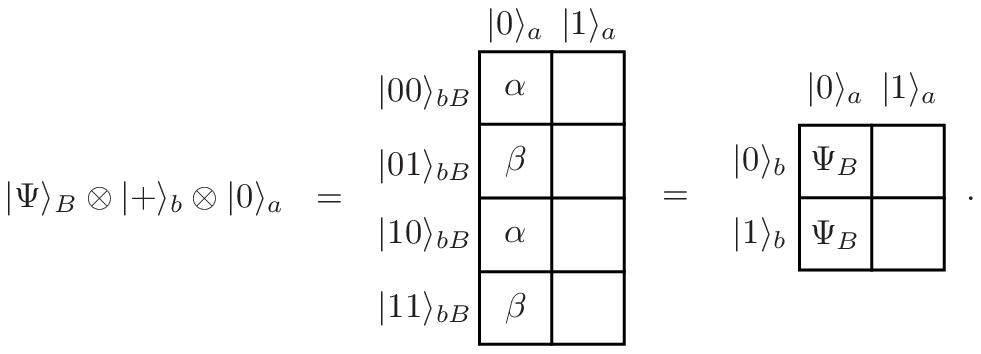}
\end{equation}
We see that it is still possible to think of there being two images. However, under these circumstances, Alice's view of the lower image of $\Psi_B$ is ``obstructed" by the presence of the upper image; the two images effectively appear as one to her, so that she is unable to act differently on one image as compared to the other. That is, anything she does to the upper image will also be done in precisely the same way to the lower image, and vice-versa. Indeed, by a simple change of basis on Bob's part, the diagram can be altered such that there will only be one image, and obviously, this change of basis cannot effect what Alice can and cannot do. More generally, if the two parties share a state of Schmidt rank $r$ on systems $a,b$ ($r=1$ for a product state; $r>1$ means the state is entangled), then they can always choose bases such that the diagram will have $r$ images appearing in diagonal blocks. Furthermore, no matter their choice of bases, $r$ is the minimum number of images they can have. 

\begin{figure}
\includegraphics{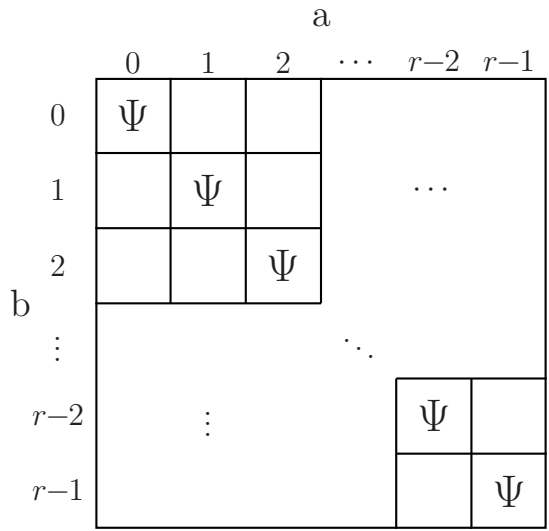}
\caption{\label{fig:manyImages}Multiple images of the state $|\Psi\rangle$ produced by Alice and Bob sharing an entangled state of Schmidt rank $r$ on systems $a,b$. Note that $|\Psi\rangle$ may itself be entangled on systems $A,B$.}
\end{figure}

We will see that it is Alice's (and Bob's) ability to ``see" the two images separately, and their resulting ability to manipulate the images \textit{differently}, that is crucial to their success in distinguishing unextendible product bases by means of our protocols. The foregoing discussion should make clear that it is the presence of entanglement between systems \textit{a,b} that provides the parties with this ability. With an entangled state and appropriate choice of bases, the images will be arranged along diagonal blocks, leaving both parties with the ability to see all the images separately from the others. This is illustrated in Fig.~\ref{fig:manyImages}.
As an example of one outcome of a measurement, Bob could perform the ``controlled-projection" on $bB$,
\begin{equation}\label{cp}
	{\cal P} = \sum_{k=0}^{r-1} |k\rangle_b\langle k|\otimes P_k,
\end{equation}
with $P_k$ a set of projectors on the Hilbert space of system $B$. If $P_kP_l=\delta_{kl}P_k$ and $\sum_kP_k=I_B$, then this outcome preserves the complete state $|\Psi\rangle$ (indeed, the complete Hilbert space of system $B$), but it is preserved in pieces distributed across the various images, and these pieces are still arranged along the diagonal of the box diagram. We will use this type of operation often in what follows.

It is worth noting the difference between maximally and partially entangled states from the perspective of these diagrams. All bipartite entangled states of Schmidt rank $r$ may be written as
\begin{equation}
	|\Phi\rangle_{ab} = \sum_{k=0}^{r-1} \lambda_k|k\rangle_a |k\rangle_b,
\end{equation}
where $\lambda_k\ne0$ are the Schmidt coefficients. With this state, each image of $|\Psi\rangle$ appearing in Fig.~\ref{fig:manyImages} will be multiplied by its corresponding $\lambda_k$. This means that the various images are inequivalent, as they are scaled differently from each other, unless all the $\lambda_k$ are the same; that is, unless $|\Phi\rangle$ is maximally entangled. When the $\lambda_k$ are not all the same then, for example, the controlled-projection of Eq.~\ref{cp} will not preserve the state unchanged, as the pieces preserved from the different images will be multiplied by different scaling factors. We will return to this point below.

In the next sections, we turn to our protocols for locally distinguishing unextendible product bases. As part of this process, Bob will perform a joint operation on systems $b,B$. By way of illustration before we proceed to the actual protocols, suppose he performs a joint measurement in the Bell basis on the state represented in Eq.~(\ref{B0SA}), obtaining outcome $\vert B_0\rangle_{bB}$. Then,

\begin{equation}\label{fig:teleport}
\includegraphics{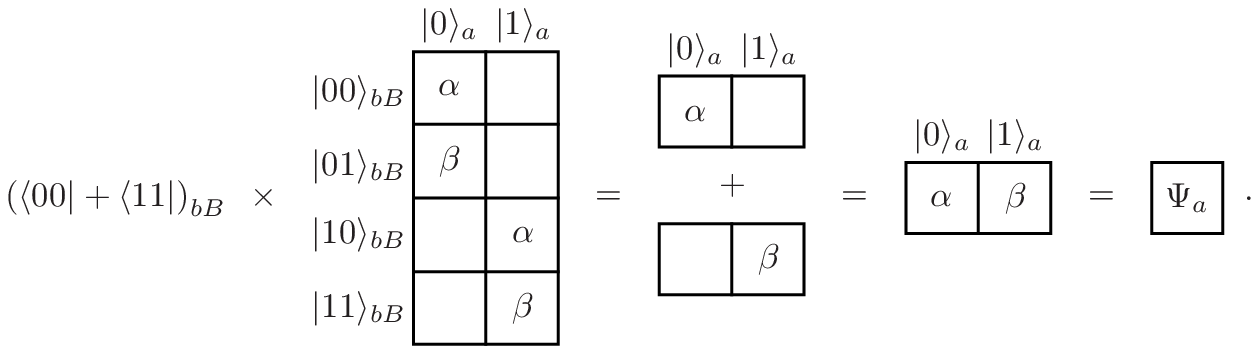}
\end{equation}
We see that the state $\vert\Psi\rangle$, originally on system $B$, is now on system $a$. That is to say, the parties have successfully teleported the state from Bob to Alice \cite{my_tele}! Notice how the $\vert 0\rangle_B$ part of $\vert\Psi\rangle_B$ is preserved from the upper image, while it is the $\vert 1\rangle_B$ part that is preserved from the lower. This is an example of what we mean by the ability to manipulate the two images differently.

In the sequel, it will be important to keep in mind that the parties cannot act on parts of rows or columns, but must act either on \textit{whole} columns or \textit{whole} rows. In particular, when Alice measures, she cannot divide a column into pieces, but must preserve whole columns, either individually, as a collection of columns, or as superpositions of them. Similarly, Bob preserves whole rows.

\section{Unextendible Product Bases on $3\otimes3$}\label{sec:3by3}
In this section, we consider some of the simplest UPB's, those on bipartite systems where each party's Hilbert space is three-dimensional. We will first look at the Tiles UPB \cite{IBM_CMP}, which is intimately related to the nine product states discovered by Bennett, et.al., with which they first demonstrated the phenomenon of nonlocality without entanglement \cite{Bennett9}. This example will serve as a useful introduction to the problem of using entanglement to distinguish UPB's with LOCC, as it will illustrate the main ideas that have proven useful in other cases, to be discussed in the following sections. In part B of this section, we will show that every UPB on $3\otimes3$ can be distinguished by LOCC using a $2\otimes2$ maximally entangled state.

\subsection{A simple example: distinguishing the Tiles UPB}\label{sec:tiles}
\begin{figure}[h]
\includegraphics{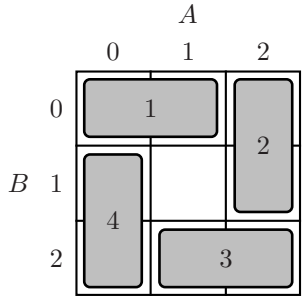}
\caption{\label{fig:tilesUPB}The Tiles UPB. The labels on the rows (columns) correspond to Bob's (Alice's) standard basis; that is, $k\rightarrow |k\rangle$.}
\end{figure}

Here, we will show how the Tiles UPB can be distinguished if the parties share an auxiliary $2\otimes2$ system in a maximally entangled state. The Tiles UPB is a set of 5 states on a $3\otimes3$ bipartite system. Omitting unimportant normalizations, these states are
\begin{eqnarray}
&|\Psi_1\rangle  = & |0-1\rangle_A|0\rangle_B,\nonumber\\
&|\Psi_2\rangle  = & |2\rangle_A|0-1\rangle_B\nonumber\\
&|\Psi_3\rangle  = & |1-2\rangle_A|2\rangle_B,\nonumber\\
&|\Psi_4\rangle  = & |0\rangle_A|1-2\rangle_B,\nonumber\\
&|F\rangle  ~= & |0+1+2\rangle_A|0+1+2\rangle_B,
\end{eqnarray}
where, for example, $|0-1\rangle$ should be understood to mean $|0\rangle-|1\rangle$. This set of states has a box-diagram representation with the simple form shown in Fig.~\ref{fig:tilesUPB}. The state $|F\rangle$, known as the stopper state, is not shown, as it would cover the whole diagram. It is well known that these five states cannot be perfectly distinguished by LOCC \cite{Bennett9} (for a very simple proof, see Appendix B of \cite{myLDPE}). Intuitively, this can be understood from the diagram by noting that in order to isolate any one of the tiles, another one of the tiles will necessarily be cut in half. For example, if Alice first projects onto $|0\rangle_A$, Bob can then isolate tile $4$, but Alice's projection will have preserved only half of tile $1$. The result of this is that the corresponding state ($|\Psi_1\rangle$ in this case) is no longer orthogonal to the stopper, so the parties will necessarily fail to distinguish for at least some cases. We see that it is the structure of the tiles relative to one another which prevents the set of states from being deterministically distinguished -- tiles $1$ and $3$ are non-orthogonal on Alice's side, forming a chain of horizontal tiles stretching completely across Alice's space and preventing her from making any measurement without destroying the orthogonality of the states in these tiles. Of course, Bob can easily pull this ``horizontal" chain apart by a measurement in his standard basis. However, since tiles $2$ and $4$ similarly form a chain of vertical tiles, they prevent him from measuring on his system. In turn, although Alice can pull the ``vertical" chain apart, she is prevented from doing so by the presence of the horizontal chain. Thus, the two chains are in some sense linked to each other -- one can say that each ``ties" the other one together -- and it is this linking that prevents the parties from distinguishing the UPB. The point is that, if the parties could find a way to unlink the two chains, then they could accomplish this feat.

\begin{figure}[h]
\includegraphics{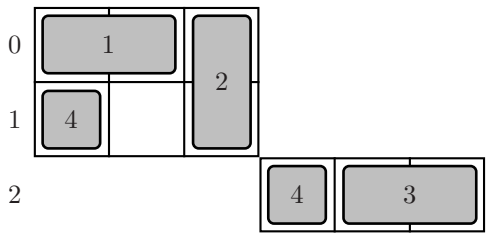}
\caption{\label{fig:shove}The Tiles UPB with the bottom row slid over to the side.}
\end{figure}

One way the two chains could be unlinked would be if they could simply slide the bottom row (labeled $2$ in Fig.~\ref{fig:tilesUPB}) over to the right. Of course, to do so would require something more than LOCC, but let us consider what this would accomplish. The diagram would then be as depicted in Fig.~\ref{fig:shove}. We see that the chain of horizontal tiles has been broken apart (tile $1$ is now orthogonal to tile $3$ on Alice's side), and it is clear that Alice can do a measurement that either isolates or excludes tile $2$, breaking apart the vertical chain.

We will soon see that, starting from the situation of Fig.~\ref{fig:shove}, it is easy for the parties to distinguish the set. Before we proceed, however, we should consider whether they can actually slide the row over to create this situation in the first place. The answer is yes, but only with additional resources. First, Alice will need a bigger Hilbert space, which could be obtained by her bringing in an ancillary qubit, system $a$. Then, the three columns on the left of Fig.~\ref{fig:shove} (the $2$-by-$3$ block) will correspond to the $|0\rangle_a$ state of this ancilla, the other three columns to the $|1\rangle_a$ state. Now we notice that in sliding the row over, the states $|\Psi_4\rangle$ and $|F\rangle$ have become entangled between systems $a$ and $B$. For example, $|\Psi_4\rangle$ has now become $|0\rangle_A(|0\rangle_a|1\rangle_B-|1\rangle_a|2\rangle_B)$, which has an entanglement of $1$ ebit. Since it previously had zero entanglement, and since entanglement cannot be increased by LOCC (on average), we see that in order for the parties to implement this protocol by LOCC, they will need to start out with $1$ ebit as an additional resource, conveniently supplied in the form of a two-qubit maximally entangled state.

\begin{figure}[h]
\includegraphics{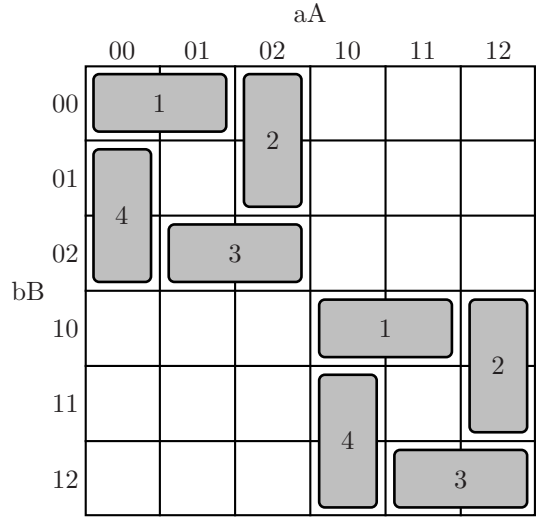}
\caption{\label{fig:tilesX2}The Tiles UPB when Alice and Bob share a $2\otimes2$ MES.}
\end{figure}

Recalling the diagrammatic presentation of the previous section, ``sliding the row over" is precisely what was effectively done in the teleportation protocol (see the center figure in Eq.~(\ref{fig:teleport})). Therefore, we already know how to do this. First of all, two qubits in the entangled state $|B_0\rangle_{ab}$, shared by Alice and Bob, create two images of their overall Hilbert space, as shown in  Fig.~\ref{fig:tilesX2}. 
Suppose that, starting from the situation depicted in this figure, Bob performs a two-outcome measurement, each outcome corresponding to rank-$3$ projectors,
\begin{eqnarray}
	B_1&=&|00\rangle_{bB}\langle00|+|01\rangle_{bB}\langle01|+|12\rangle_{bB}\langle12|,\nonumber\\
	B_2&=&|10\rangle_{bB}\langle10|+|11\rangle_{bB}\langle11|+|02\rangle_{bB}\langle02|.
\end{eqnarray}
Then, the picture of Fig.~\ref{fig:shove} is obtained with outcome $B_1$ (the middle three rows in Fig.~(\ref{fig:tilesX2}) are annihilated leaving only the first two rows and the last one), whereas $B_2$ creates a picture that differs from this only by permutation of rows and columns (the latter can therefore be handled using the exact same method as described below for $B_1$). 

\begin{figure}[h]
\includegraphics{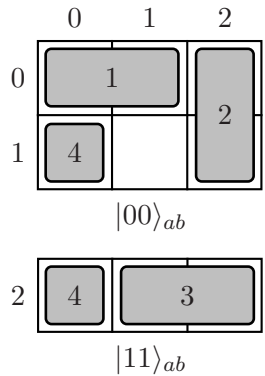}
\caption{\label{fig:UNshove}Another representation of the Tiles UPB with the bottom row slid over to the side. The numbers along the side represent the state (in the standard basis) of system $B$; those along the top are for system $A$.}
\end{figure}

To be precise, the result of bringing in the ancillary systems in state $|B_0\rangle_{ab}$, and then operating with $B_1$ on systems $bB$, is that each of the initial states is transformed as
\begin{eqnarray}\label{eqn:B1}
&|\Psi_1\rangle  \rightarrow & |0-1\rangle_A\otimes|0\rangle_B\otimes|00\rangle_{ab},\nonumber\\
&|\Psi_2\rangle  \rightarrow & |2\rangle_A\otimes|0-1\rangle_B\otimes|00\rangle_{ab},\nonumber\\
&|\Psi_3\rangle  \rightarrow & |1-2\rangle_A\otimes|2\rangle_B\otimes|11\rangle_{ab},\nonumber\\
&|\Psi_4\rangle  \rightarrow & |01\rangle_{AB}\otimes|00\rangle_{ab}-|02\rangle_{AB}\otimes|11\rangle_{ab},\nonumber\\
&|F\rangle  ~\rightarrow & |0+1+2\rangle_A\otimes\left(|0+1\rangle_B\otimes|00\rangle_{ab}+|2\rangle_B\otimes|11\rangle_{ab}\right).
\end{eqnarray}
As an explicit example, consider $|\Psi_4\rangle$. We have,
\begin{eqnarray}
	B_1(|\Psi_4\rangle_{AB}\otimes|B_0\rangle_{ab})  & = & [|00\rangle_{bB}\langle00|+|01\rangle_{bB}\langle01|+|12\rangle_{bB}\langle12|]\times[|0\rangle_A|1-2\rangle_B\otimes(|00\rangle_{ab}+|11\rangle_{ab})]\nonumber\\
				&=& |01\rangle_{bB}\otimes|00\rangle_{aA}-|12\rangle_{bB}|10\rangle_{aA}\nonumber\\
				&=& |01\rangle_{AB}\otimes|00\rangle_{ab}-|02\rangle_{AB}\otimes|11\rangle_{ab}.
\end{eqnarray}
This is represented in Fig.~\ref{fig:UNshove} by the tile $4$, which is broken up into two pieces, one associated with $|00\rangle_{ab}$ and the other with $|11\rangle_{ab}$. Similarly, the part of $|F\rangle$ that lies in this tile is also broken up in the same way. Indeed, this entire figure corresponds precisely to what results from $B_1$ acting on the original states, the resulting states having been listed above on the right in Eq.~(\ref{eqn:B1})  (note that Fig.~\ref{fig:UNshove} is equivalent to Fig.~\ref{fig:shove} except that the two pieces that have been slid across each other are now shown one above the other, the horizontal separation implied by the labels $|00\rangle_{ab}$ and $|11\rangle_{ab}$ -- the purpose of this is to save space, which will be particularly helpful in subsequent sections).

Let us now describe how the parties can proceed from here to distinguish the states. Alice makes a $4$-outcome projective measurement, and we begin by considering the first outcome, $A_1=|0\rangle_{a}\langle0|\otimes|2\rangle_{A}\langle2|$. Due to the $|0\rangle_a\langle0|$ part of this projector, the bottom row of Fig.~\ref{fig:UNshove} (that associated with $|11\rangle_{ab}$) is annihilated, while the $|2\rangle_A\langle2|$ part annihilates the left two columns ($|0\rangle_A$ and $|1\rangle_A$) of this figure. Thus, this outcome preserves tile $2$ and nothing else, so the state is either $|\Psi_2\rangle$ or $|F\rangle$. These two states have now become
\begin{eqnarray}\label{eqn:A1}
&|\Psi_2\rangle  \rightarrow & |2\rangle_A\otimes|0-1\rangle_B\otimes|00\rangle_{ab},\nonumber\\
&|F\rangle  ~\rightarrow & |2\rangle_A\otimes|0+1\rangle_B\otimes|00\rangle_{ab},
\end{eqnarray}
which can be seen either directly from what we just described as having happened to Fig.~\ref{fig:UNshove}, or by considering the action of $A_1$ on the states on the right of Eq.~(\ref{eqn:B1}). At this point, the ancillas $a,b$ can be discarded, and Bob can easily distinguish between these two remaining states by projectors onto $|0\pm1\rangle_B$.

Alice's second outcome is $A_2= |1\rangle_a\langle1|\otimes|1-2\rangle_A\langle1-2|$. The $|1\rangle_a\langle1|$ part annihilates the top two rows of Fig.~\ref{fig:UNshove}, and the $|1-2\rangle_A\langle1-2|$ part annihilates both $|\Psi_4\rangle$ (which only has support on $|0\rangle_A$) and $|F\rangle$ (whose support on the subspace spanned by $\{|1\rangle_A,|2\rangle_A\}$ is in the form $|1+2\rangle_A$). The only remaining possiblity is $|\Psi_3\rangle$, which has thus been successfully identified. Using exactly the same argument, one sees that the third outcome, $A_3= |1\rangle_a\langle1|\otimes|1+2\rangle_A\langle1+2|$ identifies $|F\rangle$.

\begin{figure}[h]
\includegraphics{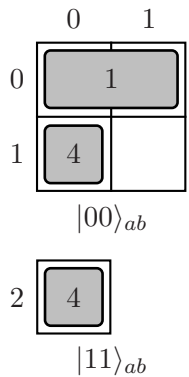}
\caption{\label{fig:A4}The result of projector $A_4$ of Eq.~(\ref{eq:A4}) acting on the states represented in Fig.~\ref{fig:UNshove}.}
\end{figure}

Alice's last outcome is a rank-$3$ projector onto the remaining part of Alice's Hilbert space, 
\begin{equation}\label{eq:A4}
	A_4=|0\rangle_{a}\langle0|\otimes(|0\rangle_{A}\langle0|+|1\rangle_{A}\langle1|)+|1\rangle_{a}\langle1|\otimes|0\rangle_{A}\langle0|.
\end{equation}
This annihilates $|\Psi_2\rangle$ and $|\Psi_3\rangle$, leaves $|\Psi_1\rangle$ and $|\Psi_4\rangle$ unchanged from what they had become in Eq.~(\ref{eqn:B1}), and now
\begin{eqnarray}
&|F\rangle\rightarrow & |0+1\rangle_A\otimes|0+1\rangle_B\otimes|00\rangle_{ab}+|0\rangle_A\otimes|2\rangle_B\otimes|11\rangle_{ab}.
\end{eqnarray}
These remaining states are represented in Fig.~\ref{fig:A4}.
Bob now follows with a two-outcome measurement, the first outcome being a projector onto $|00\rangle_{bB}$ and the second projecting onto everything else. The first outcome isolates tile $1$, so the state is either $|\Psi_1\rangle$ or $|F\rangle$, where these states have become
\begin{eqnarray}
&|\Psi_1\rangle  \rightarrow & |0-1\rangle_A\otimes|0\rangle_B\otimes|00\rangle_{ab},\nonumber\\
&|F\rangle\rightarrow & |0+1\rangle_A\otimes|0\rangle_B\otimes|00\rangle_{ab}.
\end{eqnarray}
Once again they can discard the ancillas, and Alice can distinguish with projectors onto $|0\pm1\rangle_A$. If, on the other hand, Bob obtained ``everything else", the only remaining states are now
\begin{eqnarray}
&|\Psi_4\rangle  \rightarrow & |0\rangle_A\otimes(|1\rangle_B\otimes|00\rangle_{ab}-|2\rangle_B\otimes|11\rangle_{ab}),\nonumber\\
&|F\rangle\rightarrow & |0+1\rangle_A\otimes|1\rangle_B\otimes|00\rangle_{ab}+|0\rangle_A\otimes|2\rangle_B\otimes|11\rangle_{ab}.
\end{eqnarray}
The fact that these two states are entangled makes it not quite so simple to distinguish between them. Though we know from \cite{WalgateHardy} that any two states can be distinguished by LOCC, let us finish what we have started. One party will now need to make an entangled measurement; suppose it is Bob, who can project onto $|01\rangle_{bB}\pm|12\rangle_{bB}$, with all other outcomes (however he wishes to design the rest of his projective measurement) having vanishing probability of occurring. These two outcomes leave the states as
\begin{eqnarray}
&|\Psi_4\rangle  \rightarrow & |0\rangle_A\otimes(|01\rangle_{bB}\pm|12\rangle_{bB})\otimes(|0\mp1\rangle_a),\nonumber\\
&|F\rangle\rightarrow & (|0+1\rangle_A\otimes|0\rangle_a\pm|0\rangle_A\otimes|1\rangle_a)\otimes(|01\rangle_{bB}\pm|12\rangle_{bB}).
\end{eqnarray}
These are orthogonal on Alice's side, who can measure in the standard basis of system $A$, identifying $|F\rangle$ with outcome $|1\rangle_A$. If, on the other hand, she gets outcome $|0\rangle_A$, she can complete the protocol by a measurement on her ancilla in the $|0\pm1\rangle_a$ basis.

Thus, by viewing entanglement as providing two images of the original Hilbert space that can be acted on differently by each of the parties, we have succeeded in designing a protocol that perfectly distinguishes the Tiles UPB using LOCC with an additional resource of a two-qubit maximally entangled state.

Is this amount of entanglement necessary to distinguish the Tiles UPB? It is necessary for the above protocol, as the following discussion shows. It is true that a partially entangled state, $|\Phi\rangle_{ab} = \lambda_0|00\rangle_{ab}+\lambda_1|11\rangle_{ab}$, will also provide the two images needed to slide the row over to the right. However, in this case, we must pay attention to the scaling factors, $\lambda_0$ and $\lambda_1$. If these Schmidt coefficients are unequal, as is the case for a partially entangled state, then the two images will be scaled differently, one being multiplied by $\lambda_0$, the other by $\lambda_1$. This implies that the two halves of tile $4$, the tile that is divided into two parts by sliding the row over, will be multiplied by different factors. That is,
\begin{eqnarray}
	|\Psi_4\rangle &\rightarrow& \lambda_0|00\rangle_{aA}|01\rangle_{bB}-\lambda_1|10\rangle_{aA}|12\rangle_{bB},\nonumber\\
	|F\rangle &\rightarrow& \lambda_0|00\rangle_{aA}|01\rangle_{bB}+\lambda_1|10\rangle_{aA}|12\rangle_{bB} + \cdots.
\end{eqnarray}
These two states are only orthogonal if $\lambda_0=\lambda_1$ (the ellipses in the latter represents terms that are each orthogonal to $|\Psi_4\rangle$), which means that $1$ ebit is necessary for this protocol. We see no simple way to argue that this must be the case for every successful protocol, however. One might, for example, attempt to show that at some stage in any protocol it must be that $1$ ebit is created, but we have not succeeded in doing so. Hence, it remains an open question whether it is possible to distinguish these states with less than one ebit of entanglement. We suspect that the answer to this question is no.

\subsection{All UPB'\lowercase{s} on $3\otimes3$}
\begin{figure}[h]
\includegraphics{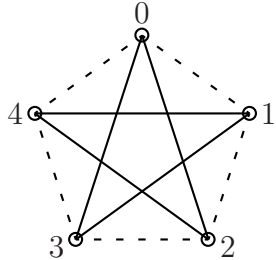}
\caption{\label{fig:ortho}The orthogonality graph of every UPB on $3\otimes3$. Each node of the graph represents one of the states in the UPB ($|\Psi_j\rangle$ is labeled as $j$), and the lines connecting two nodes indicate on which party the two nodes are orthogonal to each other: dashed lines indicate orthogonality on system $A$; solid lines, on system $B$.}
\end{figure}

We will now prove a very general theorem, applying to every UPB on $3\otimes3$. To do so, we will use two facts proven in \cite{IBM_CMP}: every such UPB (1) has exactly five members, and (2) conforms to the orthogonality graph shown in Fig.~\ref{fig:ortho}. This graph indicates which of the states in the UPB, represented in the diagram by vertices, are orthogonal to each other on Alice's side (these states are connected by dashed lines), and which on Bob's (solid lines). Writing $|\Psi_j\rangle=|a_j\rangle_A|b_j\rangle_B$ (we will assume that each of the kets appearing in this equation is normalized), then the orthogonality graph indicates, for example, that $|a_2\rangle$ is orthogonal to both $|a_1\rangle$ and $|a_3\rangle$, while $|b_2\rangle$ is orthogonal to both $|b_0\rangle$ and $|b_4\rangle$.

\begin{figure}[h]
\includegraphics{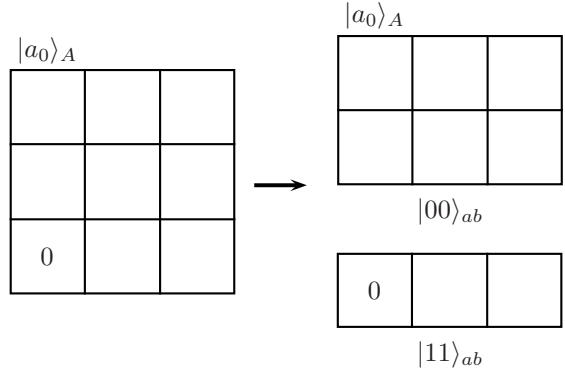}
\caption{\label{fig:3x3box}The most general UPB on $3\otimes3$, and what happens after the bottom row has been slid to the right.}
\end{figure}

We now state our theorem.
\begin{3by3} \label{3by3} One ebit of entanglement is sufficient to distinguish any UPB on $3\otimes3$, using only LOCC.
\end{3by3}
Proof:  The box diagram representing the most general UPB on $3\otimes3$ is shown on the left-hand side of Fig.~\ref{fig:3x3box}.
The reader will note the rather rudimentary nature of this representation, as I have only indicated where $|\Psi_0\rangle$ resides. The other states are not shown explicitly in order to keep the discussion completely general. Introducing a $2\otimes2$ system in the state $|B_0\rangle_{ab}$ produces two images of the overall Hilbert space, from which the parties may ``slide" one row to the right, as was done previously. If Bob slides the third row, the result is the situation represented on the right-hand side of Fig.~\ref{fig:3x3box}. Then, Alice can perform a two-outcome projective measurement, with one outcome a projection onto $|1a_0\rangle_{aA}$ (the state $|1\rangle_a$ indicates this projection is onto the part of their space that was slid to the right, represented as the bottom row on the right-hand side of the figure) which immediately identifies the state as $|\Psi_0\rangle$. Her second outcome projects onto the remaining space and excludes $|\Psi_0\rangle$, after which Alice can put the box back together by a measurement on her ancilla, projecting onto $|0+1\rangle_a$ or $|0-1\rangle_a$. If it is the latter, Bob can do a phase operation to remove the extra minus sign (this is actually unnecessary, but for the sake of discussion, it makes clear that the remainder of the protocol will not depend on which of these outcomes was obtained).  Bob can also follow with the same measurement on his ancilla, projecting onto either $|0+1\rangle_b$ or $|0-1\rangle_b$, followed by another phase operation, after which both ancillas ($a,b$) can be discarded, and to simplify the notation we will omit the subscripts labeling the kets. In any case they will be left with the box diagram on the left of Fig.~\ref{fig:3x3box}, except that $|\Psi_0\rangle$ is no longer present. That is, they are now dealing with exactly the original set of states in the original Hilbert space, but with one less state to distinguish, $|\Psi_0\rangle$ having been excluded.

Next consider what will happen when Alice does a measurement with projectors $\{|a_2\rangle\langle a_2|,I_A-|a_2\rangle\langle a_2|\}$. Since $\{|a_1\rangle,|a_3\rangle\}$ are orthogonal to $|a_2\rangle$, the first outcome of this measurement excludes $|\Psi_1\rangle$ and $|\Psi_3\rangle$, leaving only $|\Psi_2\rangle$ or $|\Psi_4\rangle$. Note that the only thing Bob has done so far is to slide the row to the right, which does not alter any of the states on his side, implying that $|b_2\rangle$ and $|b_4\rangle$ remain orthogonal to each other. Therefore, Bob can follow Alice's projection onto $|a_2\rangle$ by a projective measurement to decide between $|\Psi_2\rangle$ and $|\Psi_4\rangle$, after which they will be finished. 

If Alice obtained her second outcome, $|\Psi_2\rangle$ is excluded, leaving only $|\Psi_1\rangle$, $|\Psi_3\rangle$ and $|\Psi_4\rangle$. Again using the point noted in the previous paragraph that Bob has done essentially nothing to his space, $|b_1\rangle$ remains orthogonal to both $|b_3\rangle$ and $|b_4\rangle$. Therefore, he can now do a three-outcome measurement, projecting onto $|b_1\rangle$, $|b_3\rangle$, or the single state orthogonal to both of these. The first outcome identifies $|\Psi_1\rangle$, as it is orthogonal to the other two remaining states. The last outcome identifies $|\Psi_4\rangle$, since by construction this outcome is orthogonal to both $|b_1\rangle$ and $|b_3\rangle$. The remaining outcome excludes $|\Psi_1\rangle$, leaving it up to Alice to decide between $|\Psi_3\rangle$ and $|\Psi_4\rangle$.

Alice will be able to distinguish in this last case if and only if the states $|a_3\rangle$ and $|a_4\rangle$ have remained orthogonal to each other. Effectively, the only thing Alice has done to this point is the projection, $I_A-|a_2\rangle\langle a_2|$. So the two states on Alice's space have been transformed as
\begin{eqnarray}
	|a_3\rangle\rightarrow|a_3^\prime\rangle&=&(I_A-|a_2\rangle\langle a_2|)|a_3\rangle = |a_3\rangle-|a_2\rangle\langle a_2|a_3\rangle = |a_3\rangle,\nonumber\\
	|a_4\rangle\rightarrow|a_4^\prime\rangle&=&(I_A-|a_2\rangle\langle a_2|)|a_4\rangle = |a_4\rangle-|a_2\rangle\langle a_2|a_4\rangle,
\end{eqnarray}
given that $|a_3\rangle$ is orthogonal to $|a_2\rangle$. Since $|a_3\rangle$ is also orthogonal to $|a_4\rangle$, it is clear that $\langle a_4^\prime|a_3^\prime\rangle=0$. Hence, Alice can distinguish between the last two states, and we have proved that a $2\otimes2$ MES is sufficient to distinguish any UPB on $3\otimes3$.\hspace{\stretch{1}}$\blacksquare$

An important observation arising from this proof is that once a single state is removed, then the entangled resource is no longer needed and may be discarded. We may expect that this will not generally be the case in higher dimensions, as is implied by the discussion in the following section. Let us now turn to these more complicated UPB's.

\section{Bipartite UPB'\lowercase{s} in higher dimensions}\label{sec:GenTiles1}
\begin{figure}[h]
\includegraphics{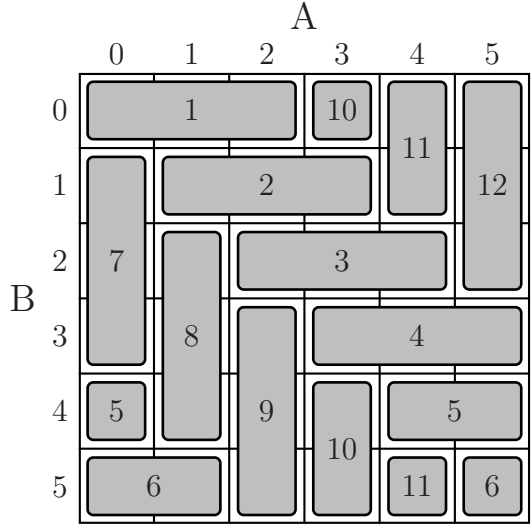}
\caption{\label{fig:GenTiles1}The GenTiles1 UPB on a $6\otimes6$ system.}
\end{figure}

In this section, we consider a generalization of the Tiles UPB to higher dimensional systems, referred to as GenTiles1 \cite{IBM_CMP,DiVincenzoTerhal}. We will first consider a specific example, that for a $6\otimes6$ system, illustrated in Fig.~\ref{fig:GenTiles1}. The generalization to arbitrary $m\otimes m$ systems, with $m$ even, is apparent from the figure -- each horizontal tile extends over half of Alice's Hilbert space, while each vertical tile covers half of Bob's. Each tile, of length $m/2$, is filled with $m/2-1$ states, each of which covers the whole tile. In addition, as with the Tiles UPB there is a stopper which covers the whole diagram and is, of course, orthogonal to all the other product states. Detailed expressions for these states will not be important for the following discussion.
In Appendix~\ref{app:GenTiles1} we prove,
\begin{GT1}\label{GT1} An $m/2\otimes m/2$ MES is sufficient to perfectly distinguish the GenTiles1 UPB on $m\otimes m$, for any even $m\ge4$, using only LOCC (the GenTiles1 UPB exists only for even $m\ge4$).
\end{GT1}
Here, we will demonstrate this on a $6\otimes6$ system.

If we consider the tiling pattern of our $6\otimes6$ example, we see a chain of links, similar to what was observed for the Tiles UPB. There is a closed chain of horizontal tiles ($1$ to $6$), as well as one of vertical tiles ($7$ to $12$), and these two chains conspire to prevent either party from making a measurement without destroying the orthogonality of the states, in precisely the same way as was seen previously for the Tiles UPB.

\begin{figure}[h]
\includegraphics{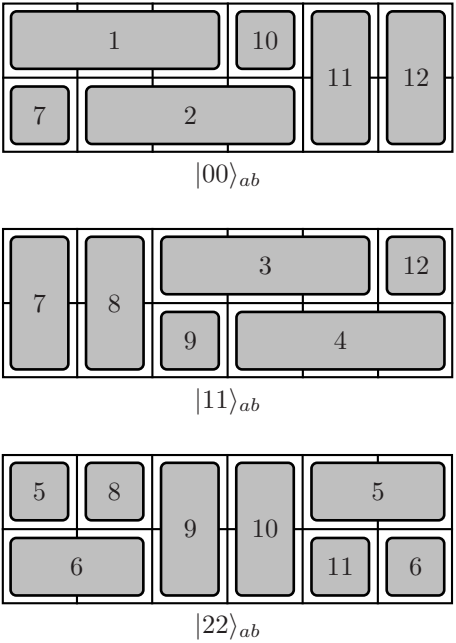}
\caption{\label{fig:GenTiles1_slide}GenTiles1 UPB after sliding each pair of rows out from under the pair above it.}
\end{figure}

\begin{figure}[h]
\includegraphics{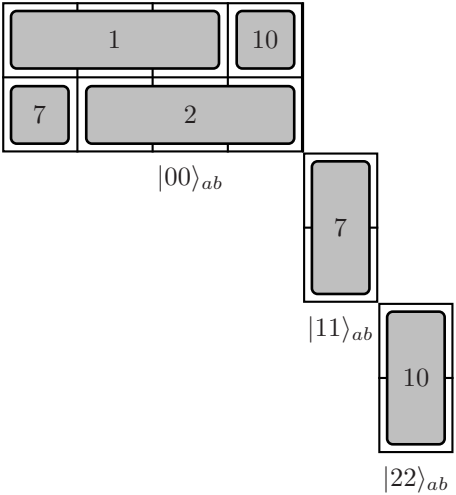}
\caption{\label{fig:GenTiles1_3}The GenTiles1 UPB after Alice's measurement.}
\end{figure}

\begin{figure}[h]
\includegraphics{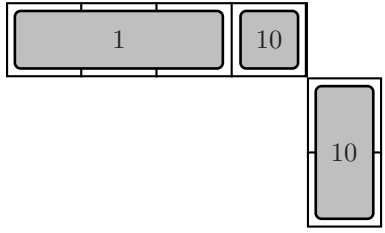}
\caption{\label{fig:GenTiles1_last}Next-to-last step in distinguishing the GenTiles1 UPB.}
\end{figure}

We can use the ideas discussed in the previous sections to demonstrate how this UPB can be distinguished with an MES of rank equal to $m/2=3$. Such an MES allows Bob to slide each successive pair of rows to the right so that each pair of rows is orthogonal on Alice's side to every other pair of rows. This is depicted in Fig.~\ref{fig:GenTiles1_slide} (recall that the labels $|kk\rangle_{ab}$, $k=0,1,2$, indicate that each pair of rows is orthogonal to the other pairs not only on Bob's side, but now also on Alice's).
From here, Alice can perform a three-outcome projective measurement which keeps only complete tiles for each outcome, dividing the set of tiles into three sets as $\{1,2,7,10\}$, $\{3,4,9,12\}$, and $\{5,6,8,11\}$. As an illustration, the first of these sets is obtained from the outcome,
\begin{eqnarray}
	A_0 = |0\rangle_a\langle0|\otimes\left(\sum_{k=0}^3|k\rangle_A\langle k|\right) + |1\rangle_a\langle1|\otimes|0\rangle_A\langle0| + |2\rangle_a\langle2|\otimes|3\rangle_A\langle3|,
\end{eqnarray}
and will leave the set of states looking like that depicted in Fig.~\ref{fig:GenTiles1_3} (the other outcomes have essentially the same form, and can be treated in the same way as this one). Bob can now do a two-outcome measurement preserving whole rows and separating the pair of tiles $\{1,10\}$ from the pair $\{2,7\}$; the first case is illustrated in Fig.~\ref{fig:GenTiles1_last}, and the other is essentially identical to this. Then, Alice can do a measurement that either distinguishes between all the states in the remaining horizontal tile, or leaves the vertical tile for Bob to distinguish. In order to assist Bob in distinguishing the vertical tile, Alice can ``put this tile back together" by a projective measurement on system $a$. For example, if it is tile $10$, she can project onto $|0+2\rangle_a,~|0-2\rangle_a$, or $|1\rangle_a$. Once tile $10$ has been isolated, the latter outcome has vanishing probability, and the other two outcomes each put this tile back into a single vertical column, after which Bob can distinguish. Thus, we have shown that for the $6\otimes6$ GenTiles1 UPB, a $3\otimes3$ MES is sufficient. The approach we have used is quite general and can be applied successfully to this UPB in any dimension, as is shown in Appendix~\ref{app:GenTiles1}.

As the ``$1$" at the end of GenTiles1 indicates, there is another UPB that generalizes Tiles, known as GenTiles2. In Appendix~\ref{app:GenTiles2}, we prove a theorem for this family of UPB's, which exists on $m\otimes n$ for any even $m\ge3$, $n>3$, and $n\ge m$. The theorem is
\begin{GT2}\label{GT2} An $\lceil m/2\rceil\otimes\lceil m/2\rceil$ MES is sufficient to perfectly distinguish the GenTiles2 UPB on $m\otimes n$ with $n\ge m$, for any dimensions $m,n$ in which it exists (excluding the case $m=3$).
\end{GT2}

To this point, we have only considered UPB's on bipartite systems. Recently, Niset and Cerf provided a construction of multipartite UPB's for (almost) any number of parties and dimensions of their Hilbert spaces \cite{NisetCerf}. In the next section, I show that every UPB obtained from their construction is LOCC-distinguishable with a single $2\otimes2$ MES shared by any pair of parties.

\section{UPB'\lowercase{s} with more than two parties}\label{sec:NisetCerf}
In this section, I consider a family of multipartite UPB's constructed by Niset and Cerf \cite{NisetCerf}.  
Their construction is quite general, producing multipartite UPB's for $N$ parties, where the $n^{th}$ party's system has arbitrary dimension $d_n$ apart from the restriction that $d_n\ge N-1$ \cite{NisetCerf}. We have found that this family of UPB's is quite weakly indistinguishable in the sense that only a relatively small amount of entanglement is needed to distinguish them. Specifically, we will prove the following theorem:
\begin{NCerf} For an arbitrary number $N$ of parties having systems of any dimensions $d_n\ge N-1$, the corresponding UPB of Niset and Cerf can be distinguished by LOCC with the aid of an ancillary $2\otimes2$ MES shared between any two parties.
\end{NCerf}
We will first prove this theorem for the specific case of $N=4$ parties, each having a $d_n=3$-dimensional system, by constructing a protocol that distinguishes this UPB. The approach we use for this example is then shown to generalize to cover all cases.

\begin{figure}[h]
\includegraphics{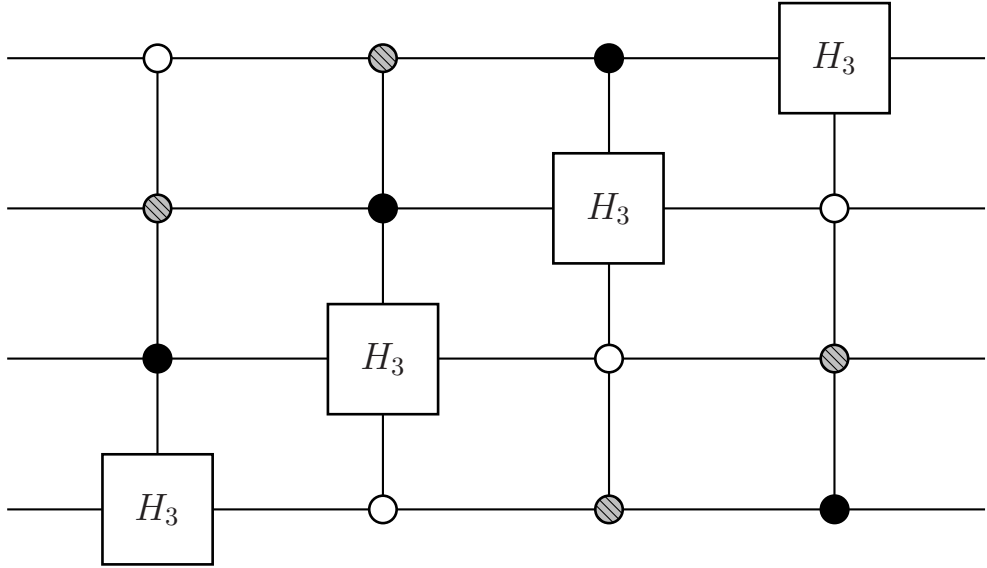}
\caption{\label{fig:NisetCerf_circuit}The Niset-Cerf construction for $4$ parties, each having a $3$-dimensional system. An open circle is ``on" if that party's state is $|0\rangle$; gray-hatched circle, if $|1\rangle$; and black circle, if $|2\rangle$. If all parties' controls are ``on", then that gate performs the DFT. Otherwise, the gate does nothing.}
\end{figure}

Let us first review the Niset-Cerf construction, using this $N=4$ and $d_n=3$ example. The states in the Niset-Cerf UPBs are products where each local state is chosen from one of two orthogonal bases: (1) the standard basis states (SB) $|0\rangle,~|1\rangle,~\cdots,~|d_n-1\rangle$; or (2) the states in the Fourier basis (FB), obtained by a discrete Fourier transform (DFT) acting on a state in SB, where the DFT is denoted below as $H_d$ for a system of dimension $d$. Though we will not need the explicit expressions of the states, which are given in \cite{NisetCerf}, it may be helpful to describe the quantum circuit that was used in the construction. The circuit of Fig.~\ref{fig:NisetCerf_circuit} illustrates the case of $N=4$ and $d_n=3$ for each $n$, with a rather obvious extension to the general case.
This circuit involves (multiply)-controlled DFT gates, and should be understood as follows: an open circle is ``on" if that party's state is $|0\rangle$; gray-hatched circle, if $|1\rangle$; and black circle, if $|2\rangle$. If all parties' controls are ``on", then that gate performs the DFT, otherwise it does nothing. It was shown that if each state of a product basis involving only SB local states is input to this circuit, the collection of output states is a complete orthogonal basis that exhibits the phenomenon of NLWE \cite{Bennett9}. They also showed that a subset of these states, when supplemented by a stopper state $|F\rangle$, is a UPB. The subset is obtained by keeping only those output states for which a DFT was performed on one of the systems (note that for any given SB input state, it is never the case that more than one system is acted on by a DFT \cite{NisetCerf}), but also omitting the states for which that DFT was performed on $|0\rangle$. The stopper is then taken as a product in which each local state is $H_d|0\rangle$. The box diagram representation of these states then includes tiles, the $n^{th}$ tile corresponding to one party and representing the $d_n-1$ states for which the DFT acted on that party's system. The stopper state covers the complete diagram including in the tiles, much like in the bipartite examples discussed above, and is the only state that lies outside the various tiles.

\begin{figure}[h]
\includegraphics{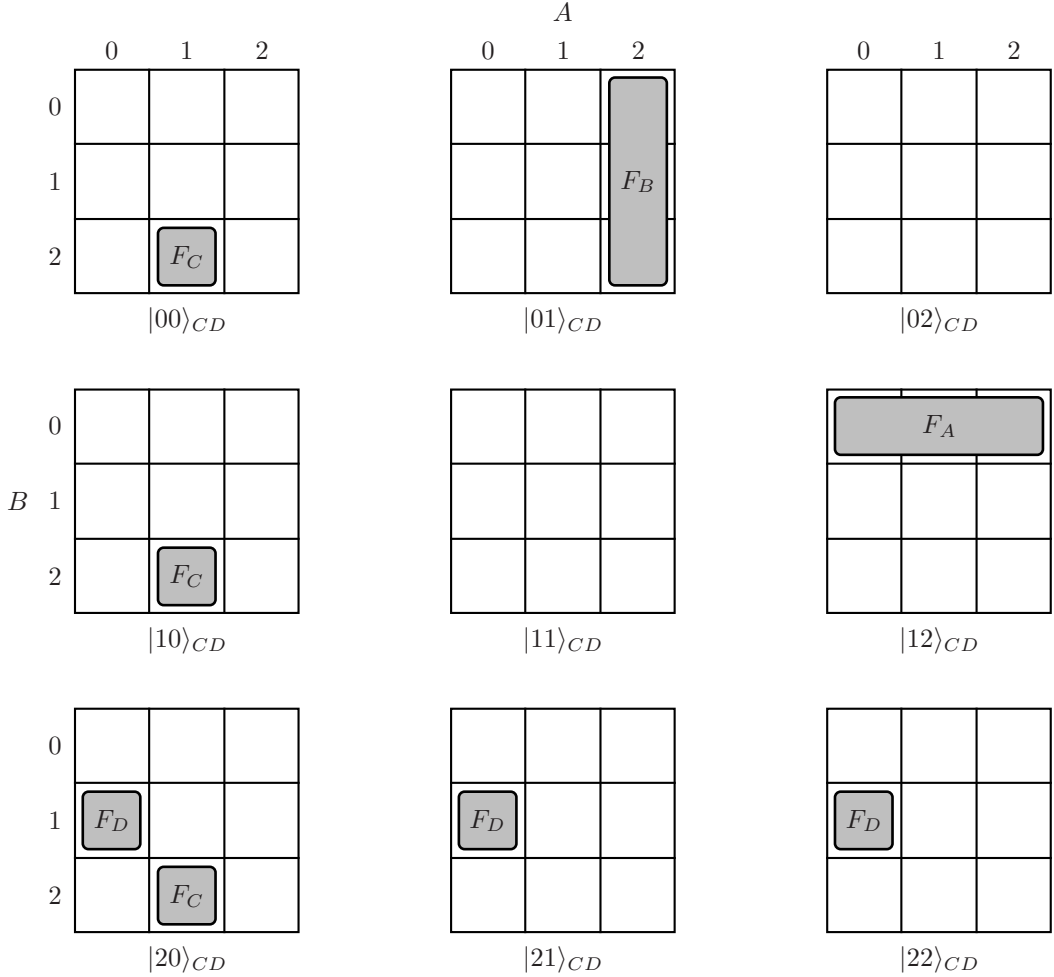}
\caption{\label{fig:NisetCerf}The Niset-Cerf construction for $4$ parties, each having a $3$-dimensional system.}
\end{figure}

This construction produces the UPB illustrated in Fig.~\ref{fig:NisetCerf}. The interpretation of this diagram as a four-dimensional box (each party's space corresponds to one of the dimensions) is as follows: each $3\otimes3$ box (I will refer to these as the ``small" boxes, as opposed to the large $4$-D box represented by the complete figure) corresponds to a fixed SB state on parties $C$ and $D$ and the full Hilbert space of $A$ and $B$. Therefore, the nine such boxes shown represent the full $4$-partite system. One may imagine this by taking each row of small boxes and stacking them on top of each other vertically (out of the page) to create a three-dimensional box. In turn, these $3$-D boxes can be stacked along a fourth dimension to create a $4$-D box.

Imagine that one party, say Alice, does a projection onto one of her SB states, say $|1\rangle$. This will pick out the second ``column" of the $4$-D box, or in other words, the collection of second columns taken from each of the individual small boxes shown in the figure. Given this outcome, the full $F_C$ tile will have been preserved across the three small boxes on the left, so the states in that tile will maintain their mutual orthogonality. However, this outcome will also cut the $F_A$ tile from the right-hand set of small boxes, preserving only part of it and precluding the possibility of successfully distinguishing the states in that tile. If, instead, she had measured in the FB and her local part of the chosen state had been one of her SB states, then she will have lost the ability to determine which local state she has. As a result, she will be unable to furnish the other parties with any useful information about what basis they should use in their measurements, and they will be left in the same bind Alice was in when she started the protocol. By noting that an FB measurement by Alice collapses each $3\otimes3$ box down to a $3\otimes1$ box that is a linear combination of all three columns in that particular box, then if another party does such a measurement, orthogonality cannot be preserved (this can most easily be seen by drawing the set of $3\otimes1$ boxes -- then, following the second party's measurement, there will always be an overlap between different tiles, the states of which will then no longer be orthogonal to each other). The same clearly holds true if the other party measures in the SB, which would destroy orthogonality within that party's tile. These ideas, along with the equivalence of all the parties, give a qualitative understanding of why this UPB cannot be distinguished by LOCC. We note that it is once again an issue of interlinking chains. Each tile by itself can be viewed as a chain stretching across that party's entire Hilbert space, and all these chains interlink each other in a way similar to the examples discussed in previous sections, in that no party can perform a measurement without breaking that chain which stretches across their space.  Breaking a chain destroys orthogonality, which is why the states cannot be distinguished by LOCC.

The point is that given one of the states in this UPB, the parties do not know whether their local part is from the SB, or from the FB. Therefore, none of them can start a measurement protocol, as they do not know what basis to use for their measurement. That is, any basis they choose to measure in, they will in some cases have chosen wrong, in which case they will destroy the orthogonality of the states and be unable to determine which state they have. 

However, if Alice's tile were removed, then she could make a measurement in the SB basis without creating any difficulties, after which the other parties would be able to distinguish all the remaining states. Though a $2\otimes2$ MES will not quite allow this tile to be immediately removed, it does allow it to be moved out of the way, as we will now see. The approach is once again to introduce entanglement in order to slide one row over. Note that ``sliding a row" means a row of the full $4$-D box; that is, for example, sliding the first row of the $4$-D box means sliding the first row of each of the small boxes in Fig.~\ref{fig:NisetCerf}.

\begin{figure}[h]
\includegraphics{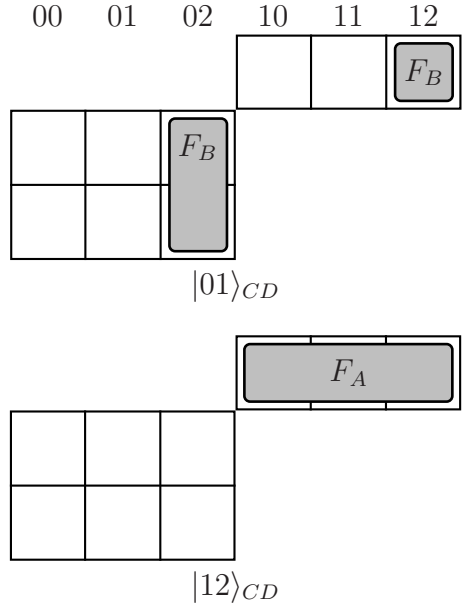}
\caption{\label{fig:NisetCerf_slide}The Niset-Cerf UPB after Bob measures and slides the top row of each small box to the right. Only the $|01\rangle_{CD}$ and $|12\rangle_{CD}$ small boxes are shown, as these are the only ones that are altered in an important way by this action. The labels across the top represent standard basis states of systems $a,A$ (for example, $00$ represents $|00\rangle_{aA}$).}
\end{figure}

Suppose Alice and Bob share a $2\otimes2$ MES, $|B_0\rangle_{ab}=|00\rangle_{ab}+|11\rangle_{ab}$. Then, referring to Fig.~\ref{fig:NisetCerf}, Bob can measure with two rank-$3$ projectors, one outcome being onto the subspace spanned by $\{|01\rangle_{bB},|02\rangle_{bB},|10\rangle_{bB}\}$ (his other outcome works similarly). Effectively, this projection slides the top row (of each small box) to the right while leaving all other rows fixed. This moves the complete tile $F_A$, and also moves only a part of tile $F_B$, as shown in Fig.~\ref{fig:NisetCerf_slide}.
At this point, Alice can make the measurement previously discussed, but instead of projecting onto $|1\rangle_A$, she projects onto $|01\rangle_{aA}$. This no longer has any effect on her tile, which has been moved out of the way. If she gets $|01\rangle_{aA}$, the only possible states are ones with Charlie's (party C) local state in the FB basis, including the stopper. This follows from the structure of the controls for the DFT gates as shown in Fig.~\ref{fig:NisetCerf_circuit}, and also can be seen from Figs.~\ref{fig:NisetCerf} and ~\ref{fig:NisetCerf_slide}. Charlie can then measure in the FB basis to determine which state it is.

If a second outcome of Alice's measurement is a projection onto $|00\rangle_{aA}$, then the same procedure works, except that now it is Dahlia (D) who measures in the FB to finish the protocol (since it is now the $F_D$ tile that has been fully preserved, all others having been excluded). Finally, a third outcome will project onto the rest of Alice's $a$-$A$ space. Then both Charlie and Dahlia know that they either have a state from their SB or their local part of the stopper. In either case, one or the other can measure in the SB to determine which of Alice or Bob has an FB state (one of them will). Note that if the chosen state is the stopper, Alice or Bob will still be able to distinguish it, even after Charlie or Dahlia have measured in the SB, since the stopper is orthogonal to all the other FB states on every local system. Thus, the party that has the FB state can measure in that basis, distinguishing the state and finishing the protocol. For example, if Charlie obtains outcome $|1\rangle_C$, then (again, this follows from the structure of the controls in Fig.~\ref{fig:NisetCerf_circuit}) Alice's system must be in an FB state. 

The procedure works for any number of parties and any dimensions of their spaces. With a single $2\otimes2$ MES shared by Alice and Bob (but by symmetry of the parties, which two of them share the MES is actually unimportant), Bob can make a two-outcome measurement for each outcome of which the $|0\rangle_B$ row of the $N$-dimensional box diagram -- the one containing Alice's complete tile -- has been slid across and separated from the rest of the diagram. Alice can again perform a $d_A$-outcome projective measurement with rank-$1$ projectors,
\begin{eqnarray}\label{Pk}
	P_k = |0\rangle_{a}\langle0\otimes|k\rangle_A\langle k|,~~0\le k\le d_A-1~~\textrm{(but~} k\ne N-1\textrm{)}
\end{eqnarray}
and one additional projector onto the remaining part of Alice's space,
\begin{eqnarray}\label{ip}
	P_\perp = I_a\otimes I_A - \sum_{k\ne N-1}^{d_A-1}P_k.
\end{eqnarray}
Note that Bob's tile has been divided by his initial measurement -- part lying in the $|00\rangle_{ab}$, and part in the $|11\rangle_{ab}$, spaces -- which is why the projector onto $|0,N-1\rangle_{aA}$ is included in the last outcome: by itself, it preserves only the part of Bob's tile that was not slid over to the right, so the states in his tile would no longer be orthogonal to each other. Alice's rank-$1$ outcomes tell her which party should measure in the FB or, if she gets outcome $P_\perp$, all parties other than Bob and herself are free to measure in the SB. One of the others measuring will determine which of Alice and Bob should then measure in the FB. In all cases, the parties succeed in distinguishing the state.

We note that if instead of distinguishing this UPB, the parties are charged with distinguishing the full NLWE basis produced by the circuit of Fig.~\ref{fig:NisetCerf_circuit} (by inputting every state in the SB of the multipartite system), then by modifying the above procedure, the parties will still succeed. This NLWE basis is obtained from the UPB by (1) removing the stopper; (2) adding a state with $H_{d_n}|0\rangle$ as the $n^{th}$ party's local part, and the other parties' local states such that this would arise from the circuit diagram of Fig.~\ref{fig:NisetCerf_circuit} (for example, input $|0\rangle_A|0\rangle_B|1\rangle_C|2\rangle_D$ to the circuit, which ``completes" tile $F_A$); and (3) then completing the basis with additional states that are all products of local SB states. In this case there are many more states to distinguish, which makes things only a bit more difficult. Bob starts with the measurement that slides the first row across, and Alice continues with the same measurement she did previously, Eqs.~(\ref{Pk}) and (\ref{ip}). Any one of the rank-1 outcomes tells her the one, and only one, party who may have an FB state (by design of her measurement, it cannot be Bob or herself). All parties other than that one can then measure in their SB, after which they will know which basis that last party should measure in to finish the protocol. If, on the other hand, she gets the projection onto the remainder of the $a$-$A$ space, only Bob or herself may have an FB state. Then, all other parties can measure in their SB, and their outcomes will determine which of Bob or Alice might have an FB state. The other of this pair then measures in the SB, after which the last party will know which basis to measure in, again completing the protocol. An important difference for this case as compared to distinguishing the UPB is that here, more measurements are required as well as more classical communication, the latter because more information must be transmitted so that subsequent parties will know what measurements they should make. This should not be surprising, however, since the parties gain more information by distinguishing the full basis, as opposed to the UPB. One might reasonably expect that to acquire more information, they will need to utilize more resources.

\section{Discussion}\label{sec:disc}
\subsection{Relationship to separable measurements}
It turns out that the UPB's discussed in this paper are all distinguishable by separable measurements (SEP). To see this, first note that all $3\otimes3$ UPB's have been shown in \cite{IBM_CMP} to be distinguishable by SEP. To show that the other UPB's considered here are distinguishable by SEP, we note they are all completable to a full (orthogonal) product basis when the stopper is removed. Then, the conclusion follows from the following theorem, which strengthens Theorem 2 of \cite{IBM_CMP}. It should be noted that the theorem of \cite{IBM_CMP} says that a UPB (any incomplete product basis, actually) is distinguishable by SEP if it is completable (perhaps in an extended Hilbert space; see \cite{IBM_CMP}) when \textit{any} single state is removed. That is to say, it must be completable no matter which state is removed in order for the conclusion to follow. Our theorem states,

\begin{SEP} A set of product states $\cal S$ is distinguishable by SEP if it is completable after a given (fixed) state, say $|\Phi_1\rangle$, is removed. \end{SEP}
\noindent Proof: The proof is quite simple. Remove $|\Phi_1\rangle$ and then add a set of states ${\cal S}_1^\perp$ to complete the basis. Design your SEP to consist of all the rank-one (product) projectors onto the states of this basis. This is a complete SEP measurement, since each projector is a product operator (and hence, separable) and their sum is the identity operator on the full (possibly extended) Hilbert space. It clearly distinguishes the states of the completed basis. It also distinguishes $\cal S$: First, each outcome identifying a state not in ${\cal S}_1^\perp$ when measuring the full basis, will identify the same state when measuring $\cal S$. Next, note that if an outcome identifies a state in ${\cal S}_1^\perp$ when measuring the full basis, then it is, by construction, orthogonal to all states in $\cal S$, with the (possible) exception of $|\Phi_1\rangle$. Hence, when measuring $\cal S$, this outcome either (1) identifies $|\Phi_1\rangle$; or (2) is orthogonal to \textit{all} states in $\cal S$ and will therefore have vanishing probability of occurring. \hspace{\stretch{1}}$\blacksquare$
\vspace{.1in}

We have attempted, without success, to apply the ideas described in this paper to devise protocols for efficiently (i.e., using less entanglement then would be required for teleportation) distinguishing other UPB's that are not known to be distinguishable by SEP. The reason for our failure may simply be that these other UPB's do not exhibit the obvious tile structure seen in the UPB's discussed here, and this tile structure clearly simplifies the task. [As for the general theorem on $3\otimes3$ UPB's, we note that it is possible to show that one can always choose local bases such that any such UPB will exhibit exactly the same tile structure seen for the Tiles UPB. This then provides an additional proof that a $2\otimes2$ MES is sufficient to distinguish any of these UPB's, as the protocol for Tiles depends only on this tile structure, not on other details of the states in the UPB.] Nonetheless, it remains an interesting open question whether or not UPB's that are distinguishable by SEP behave in a qualitatively different manner from other UPB's, with regard to the question of efficient use of entanglement to distinguish the set of states by LOCC. Might it be possible to show, for example, that all SEP operations can be implemented through LOCC by an efficient use of entanglement? Obviously, this is true for a broad subclass of SEP -- that is, the subclass of LOCC itself. But is there a close connection between non-LOCC SEP operations and LOCC itself that can be drawn through this question of efficient use of entanglement? And is there a distinction that can be made between SEP and non-SEP operations by considering the amount of entanglement that is needed to implement them by LOCC?

\subsection{UPB's and full NLWE bases}
Each of the UPB's we've considered has an associated full NLWE basis, which our protocols will also successfully distinguish. We have already described at the end of the previous section, for the multipartite Niset-Cerf construction, how to obtain the full NLWE basis from the UPB, and then how to modify the UPB-protocol so that the NLWE basis can be distinguished. The other NLWE bases can be constructed in a similar way: (1) remove the stopper; (2) add a state in each tile that is a projection of the stopper onto the subspace represented by that tile; and then (3) complete the basis with additional states (if necessary) that are all products of local SB states. For the bipartite GenTiles1 and GenTiles2, this procedure fills the whole space with ``completed" tiles (step 3 of adding SB states is not needed). Therefore, the only difference between the UPB and the NLWE is that the stopper has been replaced by several states, one in each tile. Next notice that our protocols for distinguishing the UPB's always follow the basic procedure: first, isolate a single, full tile; then second, one party distinguishes amongst the states in that tile. For each tile, then, those outcomes that identified the stopper in the case of the UPB, identified it following a projector onto that particular tile. Thus, identifying the stopper in this way is no different from identifying, in the case of the NLWE basis, the single state that was added to that tile to replace the stopper, since that single state is just the projection of the stopper onto the tile. Hence, the exact same protocol that distinguished the UPB will also distinguish the NLWE basis; there is no need for modification at all. The same conclusion holds for Tiles, except that in this case one also has to add the state $|0\rangle_A|0\rangle_B$ after removing the stopper. This SB state is easily accounted for, and the protocol for this UPB also distinguishes the NLWE, with a slight modification only needed at the end of the protocol to account for that extra $|0\rangle_A|0\rangle_{B}$ state.

\subsection{Significance of $\lceil m/2\rceil\otimes\lceil m/2\rceil$}
\begin{figure}[h]
\includegraphics{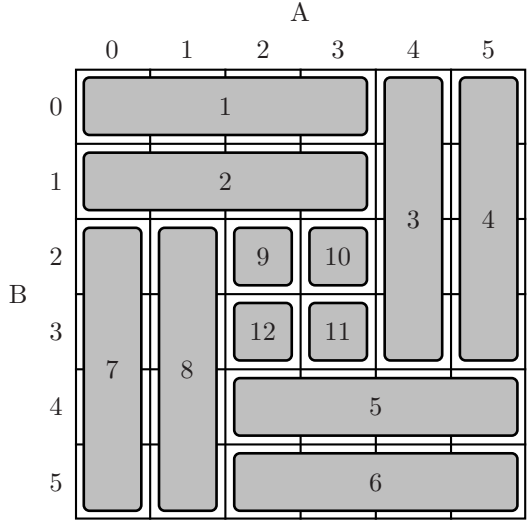}
\caption{\label{fig:Tiles^2}A complete NLWE basis on $6\otimes6$ that is distinguishable by LOCC with a $2\otimes2$ MES. Each long tile represents four orthogonal states, each of which is a superposition, with non-zero coefficients for all the SB states lying along the tile's length. The short tiles each represent a single state.}
\end{figure}

In each case we considered, it has turned out that an $\lceil m/2\rceil\otimes\lceil m/2\rceil$ MES is sufficient to distinguish the states, where $m$ is the smallest dimension of the two parties' Hilbert spaces. Is there something significant about $\lceil m/2\rceil$? We do not know the answer to this question for UPB's, but if we instead consider full NLWE bases, we can give a counter-example showing that this amount of entanglement is not always necessary. This counter-example has a structure similar to the Tiles UPB, more specifically its associated NLWE basis, the nine states with which NLWE was first demonstrated \cite{Bennett9}. The resulting basis is on $6\otimes6$ and has the tile structure shown in Fig.~\ref{fig:Tiles^2}, where each long tile represents four orthogonal states, each of which is a superposition having non-zero coefficients for all the SB states lying along the tile's length (the specific choice of such superpositions is unimportant); the short tiles each represent a single state. It can be shown, by an argument similar to that given in Appendix~B of \cite{myLDPE}, that this set of states cannot be distinguished by LOCC. Nonetheless, a $2\otimes2$ MES is sufficient to distinguish them with LOCC. One just needs to slide the bottom two rows to the right, which requires only two images of the original Hilbert space, produced by the $2\otimes2$ MES. Then a protocol similar (almost identical, actually) to that used to distinguish Tiles will also distinguish this set of states. Hence, $\lceil m/2\rceil$ is not necessary. Even so, this does not seem to imply a similar conclusion for UPB's. Although it may well be that this set can be transformed into a UPB (this does not appear to be a trivial problem), it is not at all obvious that it will be possible to distinguish the resulting UPB with a $2\otimes2$ MES.

\subsection{Tensor powers of bipartite UPB's}
Suppose one has two bipartite UPB's $S_1$ and $S_2$, given by $\{|\Psi_j^1\rangle\}$ on $m_1\otimes n_1$ and $\{|\Psi_j^2\rangle\}$ on $m_2\otimes n_2$, respectively. Then it has been shown \cite{IBM_CMP} that the set of product states , $\{|\Psi_j^1\rangle\otimes|\Psi_j^2\rangle\}$ (denoted by $S_1\otimes S_2$), is a bipartite UPB on $m_1m_2\otimes n_1n_2$. It is easy to argue that if $S_1$ can be distinguished by LOCC with an $r_1\otimes r_1$ MES, and $S_2$ with an $r_2\otimes r_2$ MES, then $S_1\otimes S_2$ is distinguishable by LOCC with a $r_1r_2\otimes r_1r_2$ MES. One can simply distinguish $S_1$ with a rank-$r_1$ MES, and $S_2$ with a rank-$r_2$ MES, confirming this claim.

\subsection{Necessary entanglement}
We have provided protocols for using entanglement to distinguish UPB's by LOCC, providing a non-trivial sufficient amount of entanglement. It would be of considerable interest to know if the amount of entanglement we use is also necessary, and if not, to know what amount is necessary. We do not know the answer to these questions, but can make a few relevant comments.

First, let us consider the question of using a partially entangled state (PES), rather than an MES. As discussed above, the difference this creates is that the various images produced by the entanglement ($r$ images for a rank-$r$ entangled state) are scaled differently from each other. In this case, our ``sliding" of rows and/or columns will tend to create non-orthogonality amongst the states within any tile that is broken up between different images. This leaves one with the impression that a PES will be difficult to use, and would probably require a good bit of creativity in devising a protocol. We suspect, but have no further compelling argument, that an MES will be necessary, at least for most cases, including those we have considered here.

Can the UPB's considered here be distinguished with an MES of rank smaller than $\lceil m/2\rceil$? We suspect not, but once again do not have a proof. Nonetheless, let us try to give a plausibility argument, considering the case of GenTiles2 discussed in Appendix~\ref{app:GenTiles2}. In our protocols for distinguishing this, and other, UPB's, the initial measurement preserves a single, complete copy of the overall Hilbert space, broken up across images. This has been done while keeping adjacent columns together so that each even-numbered short tile can be readily isolated from everything else; see Fig.~\ref{fig:GenTiles2_slide}. If one does this in a protocol using an MES of rank smaller than $\lceil m/2\rceil$, then at least one of the images will have to contribute three columns (or more). If these columns are chosen to be adjacent ones, so as to aid in isolating individual tiles from all others, then this will leave a set of states within a single image that cannot be distinguished by LOCC. The reason is that within any such trio of columns, there lies two closed, interlinking chains of states -- one composed of horizontal tiles, and the other of vertical ones. We have seen that such sets of states cannot be distinguished by LOCC, so one would need additional entanglement to produce multiple images in order to break apart these chains.

If instead one preserves three columns from a single image that are not all adjacent to one another, the result will be that more short tiles will be broken apart across different images, and we suspect this will only make it more difficult to proceed from there. Alternatively, one could design a protocol that preserves more than one complete copy of the overall Hilbert space at the first step, hoping to discover a more effective approach than the one used here. Though this might well prove successful, it is not at all obvious to us how one might proceed.

\section{Conclusions}
We have presented protocols that efficiently use entanglement to distinguish various families of UPB's by LOCC. To each of these UPB's corresponds a complete NLWE basis, which can also be distinguished by the same, or slightly modified, protocol. The design of these protocols arose from a new perspective on using entangled states as a resource, and this perspective has been described in detail in the text through the use of box diagrams. The important idea is that entanglement provides multiple images of Hilbert space, and that the parties can, independently of one another, act on these images in ways that differ from one image to the next. This allows a ``sliding" apart of Hilbert space such that initially non-orthogonal pairs of local states end up being orthogonal, aiding the process of distinguishing the set of states. It should be noted that our protocols do not rely on details of the individual states, but only on the general way they are distributed through the space.

An interesting point \cite{refThnx} is that in all of our protocols, it has been necessary that the parties be able to communicate (classically) back-and-forth. Indeed, it appears as though one-way CC will not be sufficient for the parties to distinguish these sets of states unless, of course, they share enough entanglement to teleport everything to a single party. However appearances can be deceiving, especially when one is dealing with quantum systems, so we simply note this as an important open question deserving of further study.

\begin{acknowledgments}
This work has been supported in part by the National Science Foundation through Grant PHY-0456951. I am very grateful for numerous helpful discussions with Bob Griffiths and others in his research group.
\end{acknowledgments}
\appendix

\section{T\lowercase{he} G\lowercase{en}T\lowercase{iles}1 UPB}\label{app:GenTiles1}
In this appendix, we show that the GenTiles1 UPB in any even dimension $m\ge4$ can be distinguished using the types of protocols described in the main text. Specifically, we will prove the theorem stated there,

\vspace{.1 in}
\noindent {\bf Theorem \ref{GT1}} \textit{An $m/2\otimes m/2$ MES is sufficient to perfectly distinguish the GenTiles1 UPB on $m\otimes m$, for any even $m\ge4$, using only LOCC} (\textit{the GenTiles1 UPB exists only for even $m\ge4$}).

\vspace{.1 in}
\noindent Proof: The cases of $m/2$ even and odd require slightly different discussions, though the approach in both cases is very similar to that given in the main part of this paper for $m=6$. The stated amount of entanglement produces $m/2$ equivalent images of the original Hilbert space. In the usual way, Bob can do a projective measurement that picks out a unique pair of rows from each image and slides each pair to the right so that no part of an individual pair lies vertically above any part of any other pair (each pair is then orthogonal to every other pair on Alice's side, as well as on Bob's). Since each pair obviously consists of two rows, and $2\times m/2=m$, he has preserved the whole Hilbert space of system $B$ essentially unchanged, but has broken it up into pieces with the help of system $b$. Alice will then begin the process of separating out subsets of the tiles with her first measurement. For the following discussion, it will be useful to recall that all tiles have length $m/2$.

\begin{figure}[h]
\includegraphics{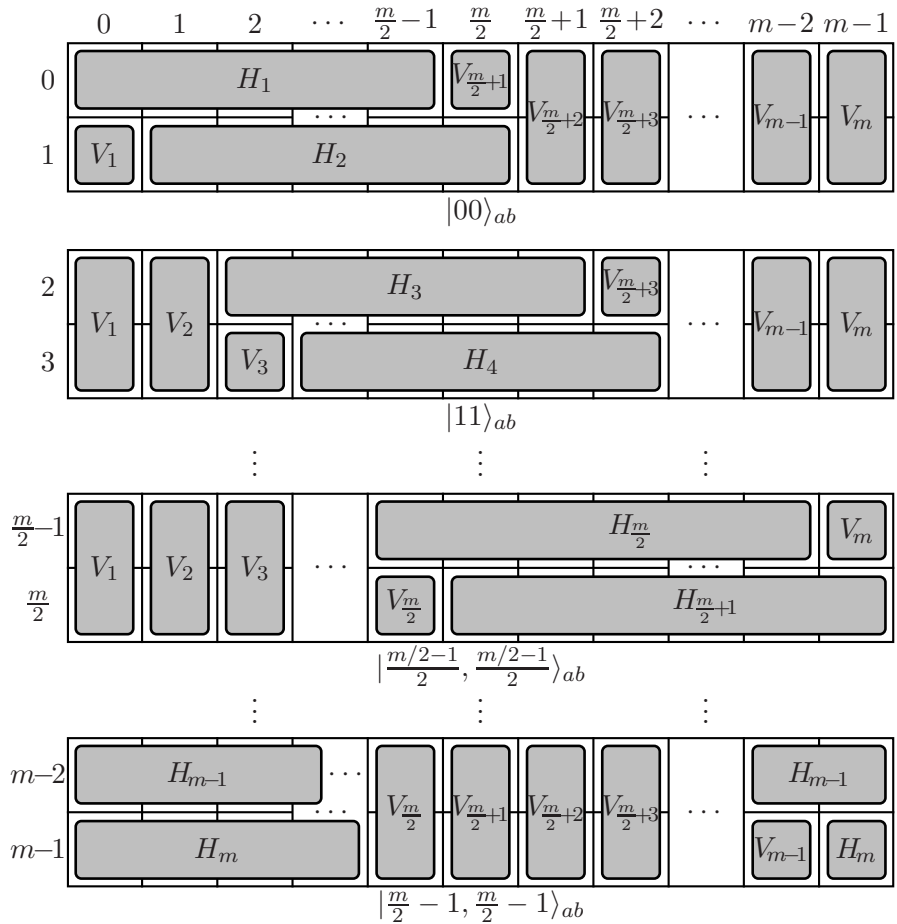}
\caption{\label{fig:gt1b0}GenTiles1 following Bob's first measurement with outcome $B_0$.}
\end{figure}

Bob's initial measurement has $m/2$ outcomes, each of which corresponds to the projectors,
\begin{equation}\label{Bl}
	B_l=\sum_{k=0}^{m/2-1}|\{k+l\}\rangle_b\langle \{k+l\}|\otimes(|2k\rangle_B\langle 2k|+|2k+1\rangle_B\langle 2k+1|),
\end{equation}
with $l$ running from $0$ to $m/2-1$, and $\{x\}$ means $x$ (mod $m/2$). Each $|\{k+l\}\rangle_b\langle \{k+l\}|$ identifies which image that term acts on, and then the associated rank-$2$ projector on the $B$ space identifies the two rows from that image that are preserved, all other rows being annihilated. We will discuss only $l=0$, the other outcomes being handled in an essentially identical way. For the case that $m/2$ is odd, which we focus on first, outcome $B_0$ leaves the situation depicted in Fig.~\ref{fig:gt1b0}.
Notice that the horizontal ($H$) tiles are preserved entirely from a single image, while the vertical ($V$) tiles are split up amongst several of the images. Many pieces of the $V$-tiles have sub-tiles of vertical length $2$, that is, they stretch across both rows preserved from that image. However, since $m/2$ is odd, and each tile has full length equal to $m/2$, each of the $V$-tiles has a single piece (from one of the images) that has vertical length of only $1$. For example, tile $V_1$ has a length-$1$ sub-tile in the uppermost pair of rows, preserved from the $|00\rangle_{ab}$ image, and then stretches downward, with several length-$2$ sub-tiles lying in each of the subsequent images down to the $|\frac{m/2-1}{2},\frac{m/2-1}{2}\rangle_{ab}$ image where the $V_1$ tile ends. At the other end of the $H$-tiles in the $|00\rangle_{ab}$ image ($H_1$ and $H_2$) lies a length-$1$ piece of the $V_{m/2+1}$ block, which stretches upward from there (with wrapping around from top of the diagram to bottom), having several length-$2$ pieces lying in other images. It is not difficult to see that the four tiles, $H_1$, $H_2$, $V_1$, and $V_{m/2+1}$, can be separated out from all the others, preserving these four tiles in their entirety. Indeed, the projector $A_0$ on Alice's space does the trick, where
\begin{eqnarray}\label{Ak}
	A_k&=&|k\rangle_a\langle k|\otimes \sum_{j=2k}^{[m/2+2k]}|j\rangle_A\langle j| \nonumber\\
		&+& \sum_{i=\{k+1\}}^{\{m/4-1/2+k\}}|i\rangle_a\langle i|\otimes |2k\rangle_A\langle 2k|\nonumber\\
		&+&\sum_{i=\{m/4+1/2+k\}}^{\{m/2+k-1\}}|i\rangle_a\langle i|\otimes |[m/2+2k]\rangle_A\langle [m/2+2k]|),
\end{eqnarray}
and $[x]\equiv x~ (\textrm{mod}~ m)$. Specifically, and referring to Fig.~\ref{fig:gt1b0}, the first line of this equation for $k=0$ preserves the first $m/2+1$ columns of the $|00\rangle_{ab}$ image, including the entire tiles $H_1$ and $H_2$ as well as the length-$1$ pieces of tiles $V_1$ and $V_{m/2+1}$. Since entire tiles must be preserved in order to avoid destroying orthogonality, we must also keep the remaining parts of these two $V$-tiles. These remaining pieces are all of length-$2$, so Alice can easily do this without the undesired effect of including pieces of other tiles: the second line in Eq.~(\ref{Ak}) preserves the rest of $V_1$, while the third line preserves the rest of $V_{m/2+1}$. Furthermore, in the same way, $A_1$ preserves $H_3$, $H_4$, $V_2$, and $V_{m/2+3}$ in their entirety, and in general $A_k$ preserves $H_{2k+1}$, $H_{2k+2}$, $V_{2k+1}$, and $V_{[m/2+2k+1]}$ (the mod $m$ in the index on the latter $V$ should be ignored if that index is equal to $m$). Hence Alice's (complete) measurement represented by the $A_k$ (with $k=0,\cdots,m/2-1$) divides the tiles into mutually exclusive sets, each set containing four tiles.

These four-tile sets are all distributed across the images in essentially the same way, so the remainder of the protocol is also essentially the same no matter which outcome, $A_k$, Alice obtained in her measurement. Let us then show how to proceed with $A_0$. So far, we have assumed that outcome $B_0$ by Bob was followed with outcome $A_0$ by Alice. The situation is still depicted by Fig.~\ref{fig:gt1b0} if one imagines that all tiles have been erased except for the four that have been preserved, $H_1$, $H_2$, $V_1$, and $V_{m/2+1}$. If Bob now repeats the measurement shown in Eq.~(\ref{Bl}), he will again obtain outcome $B_0$ with certainty, without altering anything by doing so. Let him instead refine this measurement by leaving all other $B_l$ unchanged, but splitting $B_0$ into two separate outcomes, as
\begin{eqnarray}
	B_{01}&=&|0\rangle_b\langle 0|\otimes|0\rangle_B\langle 0|+\sum_{k=m/4+1/2}^{m/2-1}|k\rangle_b\langle k|\otimes(|2k\rangle_B\langle 2k|+|2k+1\rangle_B\langle 2k+1|)\nonumber\\
	B_{02}&=&|0\rangle_b\langle 0|\otimes|1\rangle_B\langle 1|+\sum_{k=1}^{m/4-1/2}|k\rangle_b\langle k|\otimes(|2k\rangle_B\langle 2k|+|2k+1\rangle_B\langle 2k+1|).
\end{eqnarray}
He will obtain one of these two outcomes (all other outcomes now have vanishing probability); the first one preserves all of tiles $H_1$ and $V_{m/2+1}$ while the second preserves all of tiles $H_2$ and $V_1$. Whichever of these outcomes he obtains, Alice can follow with a measurement (a refinement of $A_0$) that isolates either the remaining $H$-tile or the remaining $V$-tile. If, for example, Bob obtained $B_{02}$, Alice does
\begin{eqnarray}
	A_{01}&=&|0\rangle_a\langle 0|\otimes\sum_{j=1}^{m/2}|j\rangle_A\langle j|\nonumber\\
	A_{02}&=&\sum_{k=0}^{m/4-1/2}|k\rangle_a\langle k|\otimes|0\rangle_A\langle 0|,
\end{eqnarray}
these being the only outcomes occurring with non-zero probability, the first one isolating tile $H_2$ and the second one isolating $V_1$. In any case, the states in the remaining $H$-tile can be directly distinguished by Alice, while for the $V$-tile, Alice will need to put this tile back together by a measurement in the Fourier basis on system $a$. Then, Bob can distinguish amongst the states in that tile. This completes the protocol for the case $m/2$ odd.

In the foregoing, Alice designed her outcome $A_0$ as follows: it should preserve tile $H_1$, in which case she must preserve all of this tile. To do so, she also preserves a length-$1$ piece of $V_1$ and almost all of $H_2$, so she must preserve all of these tiles, as well (recall that her measurements preserve entire columns, which looking back at Fig.~\ref{fig:gt1b0}, have vertical length of $2$). Preserving all of $H_2$ means that a length-$1$ piece of $V_{m/2+1}$ is preserved, so this tile must also be preserved in its entirety. It turns out, as shown above, that these four tiles can all be completely preserved without keeping any part of any other tile. Alice's other outcomes work the same way, which is why the protocol is able to succeed.

\begin{figure}[h]
\includegraphics{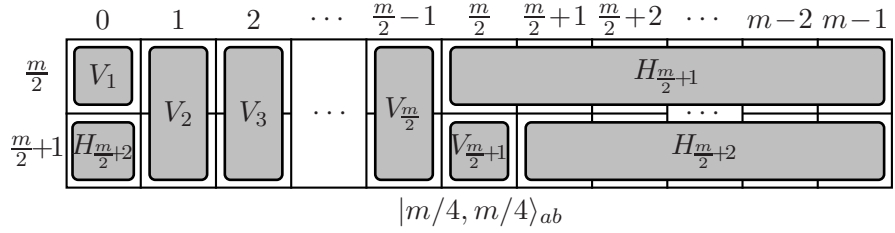}
\caption{\label{fig:gt1even}GenTiles1 following Bob's first measurement with outcome $B_0$ when $m/2$ is even. We only show image $|m/4,m/4\rangle_{ab}$, as it is the only change needed to Fig.~\ref{fig:gt1b0}, replacing image $|\frac{m/2-1}{2},\frac{m/2-1}{2}\rangle_{ab}$ there.}
\end{figure}

Now consider what is different when $m/2$ is even instead of odd. Bob's first measurement is unchanged, preserving two rows from each of the $m/2$ images provided by the $m/2\otimes m/2$ MES. The first pair of rows, from the $|00\rangle_{ab}$ image, still contains within it $H_1$ and $H_2$, along with length-$1$ pieces of $V_1$ and $V_{m/2+1}$. Alice can design $A_0$ by intending for it to preserve $H_1$, in which case this outcome must preserve all of this tile, along with $H_2$, $V_1$, and $V_{m/2+1}$. So far, all is as it was for $m/2$ odd. But preserving the rest of one of these vertical tiles will no longer mean \textit{only} keeping an additional set of length-$2$ sub-tiles, which along with the single length-$1$ sub-tile already noted, would make the total length odd. Rather, since $m/2$ is now even and is the length of each of the tiles, there must be a second length-$1$ sub-tile for each of these vertical tiles. Since the additional piece of $V_1$ (to focus on a specific $V$-tile) is length-$1$, there will be an $H$-tile lying below it within the same image, which in this case will be the $|m/4,m/4\rangle_{ab}$ image, shown in Fig.~\ref{fig:gt1even} (this should replace the $|\frac{m/2-1}{2},\frac{m/2-1}{2}\rangle_{ab}$ image in Fig.~\ref{fig:gt1b0} because now $m/4-1/2$ is not an integer). This means that in order for Alice's outcome to include that second length-$1$ piece of $V_1$, it must also preserve this $H$-tile, which is seen from Fig.~\ref{fig:gt1even} to be $H_{m/2+2}$. To preserve all of this latter tile, we see that tile $H_{m/2+1}$ must also be preserved, which implies that tile $V_{m/2+1}$ must be preserved. Here, the chain of implications, preserving one tile implying that another need be preserved, closes on itself because we already knew that $V_{m/2+1}$ had to be preserved. Therefore, the design of $A_0$ is complete, and it preserves only $6$ tiles: $H_1$, $H_2$, $H_{m/2+1}$, $H_{m/2+2}$, $V_1$, and $V_{m/2+1}$.

Further thought along these lines shows that the (incomplete -- to be completed below) measurement with operators
\begin{eqnarray}\label{A2k}
	A_{2k}&=&|k\rangle_a\langle k|\otimes \left(\sum_{j=2k}^{[m/2+2k]}|j\rangle_A\langle j|\right)+|m/4+k\rangle_a\langle m/4+k|\otimes \left(\sum_{j=[m/2+2k]}^{[m+2k]}|j\rangle_A\langle j|\right) \nonumber\\
		&+& \left(\sum_{i=k+1}^{m/4+k-1}|i\rangle_a\langle i|\right)\otimes |2k\rangle_A\langle 2k|\nonumber\\
		&+&\left(\sum_{i=\{m/4+k+1\}}^{\{m/2+k-1\}}|i\rangle_a\langle i|\right)\otimes |[m/2+2k]\rangle_A\langle [m/2+2k]|),
\end{eqnarray}
with $k=0,\cdots,m/4-1$, will preserve tiles $H_{2k+1}$, $H_{2k+2}$, $H_{m/2+2k+1}$, $H_{m/2+2k+2}$, $V_{2k+1}$, and $V_{m/2+2k+1}$. The two terms in the first line preserve all these $H$-tiles and the length-$1$ pieces of these $V$-tiles, while the next two lines preserve all the remaining length-$2$ pieces of the $V$-tiles.

Given one of these outcomes, Bob can divide the set of $6$ tiles into two, each having one $V$-tile and two $H$-tiles. For example if Alice obtained $A_0$, Bob can preserve tiles $H_1$, $H_{m/2+2}$, and $V_{m/2+1}$ in one outcome, and $H_2$, $H_{m/2+1}$, and $V_{1}$ in the other, by a measurement (again a refinement of his initial $B_0$) including
\begin{eqnarray}
	B_{01}&=&|0\rangle_b\langle 0|\otimes|0\rangle_B\langle 0|+|m/4\rangle_b\langle m/4|\otimes|m/2+1\rangle_B\langle m/2+1|\nonumber\\
		&+&\left(\sum_{k=m/4+1}^{m/2-1}|k\rangle_b\langle k|\right)\otimes(|2k\rangle_B\langle 2k|+|2k+1\rangle_B\langle 2k+1|)\nonumber\\
	B_{02}&=&|0\rangle_b\langle 0|\otimes|1\rangle_B\langle 1|+|m/4\rangle_b\langle m/4|\otimes|m/2\rangle_B\langle m/2|\nonumber\\
		&+&\left(\sum_{k=1}^{m/4-1}|k\rangle_b\langle k|\right)\otimes(|2k\rangle_B\langle 2k|+|2k+1\rangle_B\langle 2k+1|)
\end{eqnarray}
(these are the only outcomes with non-zero probability). The first line of $B_{01}$ ($B_{02}$) preserves the first (second) row of image $|00\rangle_{ab}$ and the second (first) row of image $|m/4,m/4\rangle_{ab}$, which contain the two horizontal tiles, $H_1$ and $H_{m/2+2}$ ($H_2$ and $H_{m/2+1}$), and the length-$1$ pieces of the vertical tile $V_{m/2+1}$ ($V_{1}$). The second line of these two outcomes preserves all the length-$2$ pieces of the appropriate vertical tile. Following either of these outcomes, Alice will be able to isolate individual tiles by a measurement that has only those three outcomes (one for each tile) having non-zero probability. This can be seen by noticing that the remaining vertical tile is orthogonal to both remaining horizontal tiles on system $A$, and the two horizontal tiles lie in different images, meaning they are orthogonal to each other on system $a$. Once an individual tile has been isolated, the parties can readily distinguish amongst the states in that tile.

We just need to complete Alice's measurement begun in Eq.~(\ref{A2k}). Considering Fig.~\ref{fig:gt1b0}, we see that the top of tile $V_2$ is a length-$2$ sub-tile in the $|11\rangle_{ab}$ image, which means that all $V_2$ sub-tiles lying below this will also be of length $2$. The last sub-tile of $V_2$ is that appearing in Fig.~\ref{fig:gt1even}. This means there are no length-$1$ pieces of $V_2$, all pieces having length $2$. The same conclusion will apply to all even-numbered vertical tiles, $V_{2k}$. Therefore, these tiles can be individually isolated if Alice includes in her measurement of Eq.~(\ref{A2k}) additional projectors of the form,
\begin{eqnarray}
	A_{2k+1}&=&\left(\sum_{j=k}^{m/4+k-1}|j\rangle_a\langle j|\right)\otimes|2k+1\rangle_A\langle 2k+1|,
\end{eqnarray}
and here $k$ runs from $0$ to $m/2-1$. Outcome $A_1$ isolates tile $V_2$, and generally, $A_k$ isolates tile $V_{2k}$. In a way similar to what was described above for the odd-$m/2$ case, the states in these isolated tiles can be distinguished by Bob after Alice first puts the tile back together with a measurement in the Fourier basis of system $a$.
This completes the proof of Theorem~\ref{GT1}.\hspace{\stretch{1}}$\blacksquare$

\section{T\lowercase{he} G\lowercase{en}T\lowercase{iles}2 UPB}\label{app:GenTiles2}
\begin{figure}[h]

\includegraphics{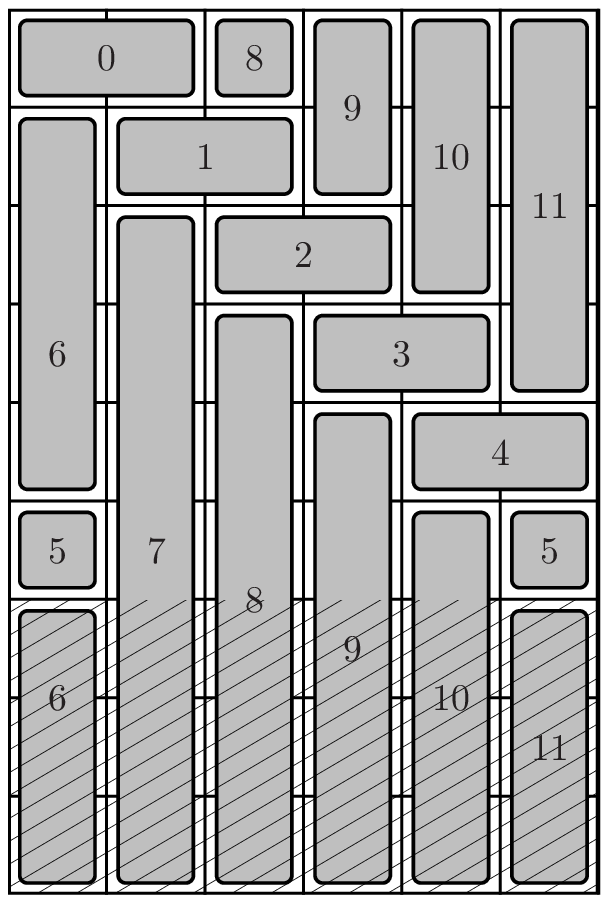}
\vspace{.5in}
\caption{\label{fig:GenTiles2_6xn}GenTiles2 for a $6\otimes n$ system. The cross-hatching in the bottom section is meant to indicate that the vertical tiles can have arbitrary lengths (Bob's space is of dimension $n$, which can, but need not, be larger than $6$).}
\end{figure}

In this appendix, we provide an LOCC protocol for distinguishing another generalization of the Tiles UPB, this one referred to as GenTiles2. The construction for this UPB is valid for any $m\otimes n$ system such that $m\ge3$, $n>3$ and $n\ge m$. As illustrated in Fig.~\ref{fig:GenTiles2_6xn} for the $6\otimes n$ case, the horizontal tiles are all two-dimensional, and the vertical tiles are $(n-2)$-dimensional (in the figure, we may imagine that the bottom, hatched rows have an arbitrary vertical extension downward -- that is, of length $n-6$). For the present purposes, we will not need the detailed expressions of the individual states, but we give them here for clarity. 
\begin{eqnarray}
	|S_{j}\rangle&=&\frac{1}{\sqrt{2}}(|j\rangle_A-|j+1\rangle_A)\otimes|j\rangle_B,\nonumber\\
	|L_{jk}\rangle&=&|j\rangle_A\otimes\frac{1}{\sqrt{n-2}}\left(\sum_{i=0}^{m-3}\omega^{ik}|i+j+1~\textrm{mod}~m\rangle_B+\sum_{i=m-2}^{n-3}\omega^{ik}|i+2\rangle_B\right),\nonumber\\
	|F\rangle&=&\frac{1}{\sqrt{mn}}\sum_{i=0}^{m-1}\sum_{j=0}^{n-1}|i\rangle_A|j\rangle_B,
\end{eqnarray}
with $\omega=\exp[{2\pi i /(n-2)}]$, $0\le j \le m-1$, and $1\le k \le n-3$ ($S$ refers to the short tiles, $L$ to the long ones, and $F$ is the stopper).

Though differing as to the detailed tiling, GenTiles2 displays a pattern similar to that of Tiles. One can view the tiles as two closed chains, one made of the horizontal tiles and the other of the vertical ones. These two chains link each other in the sense that the presence of one prevents the breaking apart of the other without destroying orthogonality. We will now prove the following theorem, also stated in the main text:

\vspace{.1 in}
\noindent {\bf Theorem \ref{GT2}} \textit{An $\lceil m/2\rceil\otimes\lceil m/2\rceil$ MES is sufficient to perfectly distinguish the GenTiles2 UPB on $m\otimes n$ with $n\ge m$, for any dimensions $m,n$ in which it exists (excluding the case $m=3$).}

\begin{figure}[h]
\includegraphics{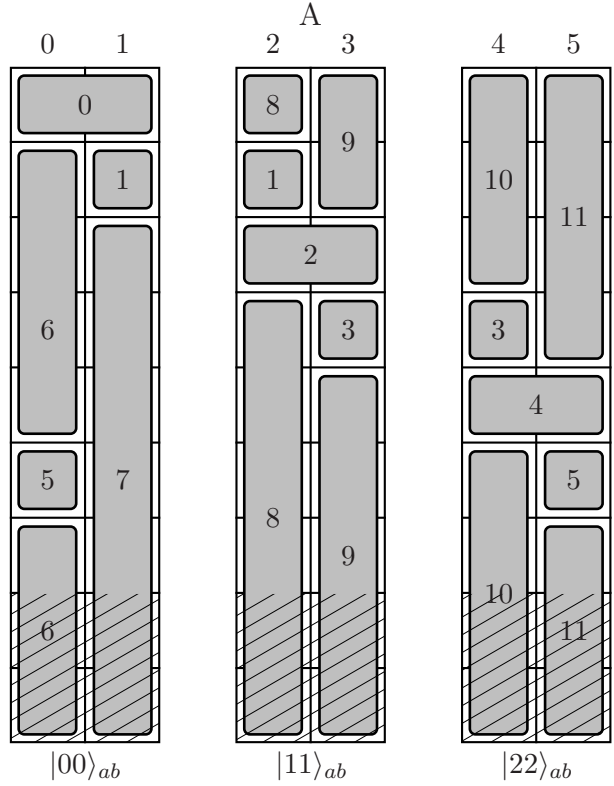}
\caption{\label{fig:GenTiles2_slide} GenTiles2 for a $6\otimes n$ system after Alice's measurement, indicating the vertical sliding of pairs of columns relative to each other.}
\end{figure}

\vspace{.1 in}
Proof: The method is similar to the approach we've used previously and works for any $m$ and $n$ (excluding $m=3$; see below). Here Alice goes first, doing a projective, entangled (between systems $A$ and $a$) measurement that picks out one pair of adjacent columns from each of the $\lceil m/2\rceil$ images of the original Hilbert space, such that each column is preserved once and only once (from a single image) in any given outcome. If $m$ is odd, then the last column will be preserved without a partner in each of these outcomes, but otherwise all columns are preserved as adjacent pairs. The result is as though each adjacent pair of columns has been slid vertically downward so that they no longer have a horizontal overlap with any part of the other pairs of columns; that is, each pair of columns is orthogonal to every other pair not only on Alice's side, but now also on Bob's (see Fig.~\ref{fig:GenTiles2_slide}).
Bob can now do a measurement that either (1) isolates a single one of the even-numbered short tiles, $\{0,2,4,\cdots\}$ (since each of these tiles lie alone in their own rows), after which Alice can distinguish between the corresponding $S_j$ and the stopper $F$; or (2) preserves everything else (other than these even-numbered short tiles). When $m$ is even, the latter outcome removes all the even-numbered tiles, so every link of the chain of horizontal tiles has been broken. Then it is a simple task for Alice to do an $m/2$-outcome measurement, each outcome being a rank-$2$ projector that preserves, in its entirety, one of the odd-numbered short tiles, along with the associated vertical tiles (those lying in the same columns). These outcomes ``shift" the pairing of the columns -- whereas before, columns $0$ and $1$, $2$ and $3$, $\cdots$, $m-2$ and $m-1$, were paired; now it will be columns $1$ and $2$, $3$ and $4$, $\cdots$, $m-1$ and $0$. Each of Alice's outcomes leaves three tiles (one horizontal tile, and two vertical ones) -- for example, tiles $1$, $7$, and $8$ are preserved together (or tiles $\{3, 9, 10\}$, or $\{5, 6, 11\}$) in Fig.~\ref{fig:GenTiles2_slide}. Recalling that the even short tiles have been removed -- specifically, in Fig.~\ref{fig:GenTiles2_slide}, tiles $0$, $2$, and $4$ are no longer present -- Bob can follow any of these outcomes of Alice's with a $2$-outcome projective measurement of his own, where one outcome isolates the remaining (odd-numbered) short tile and the other projects onto everything else. If it is the short tile, then Alice can distinguish between the two possible states, $S_j$ and $F$, whereas if it is everything else, she can isolate one or the other of the remaining vertical tiles. Once these vertical tiles are isolated, Bob can distinguish amongst the states in that tile. For $m$ even, then, this description includes all the columns, and therefore provides a complete protocol for distinguishing the states.

\begin{figure}[h]
\includegraphics{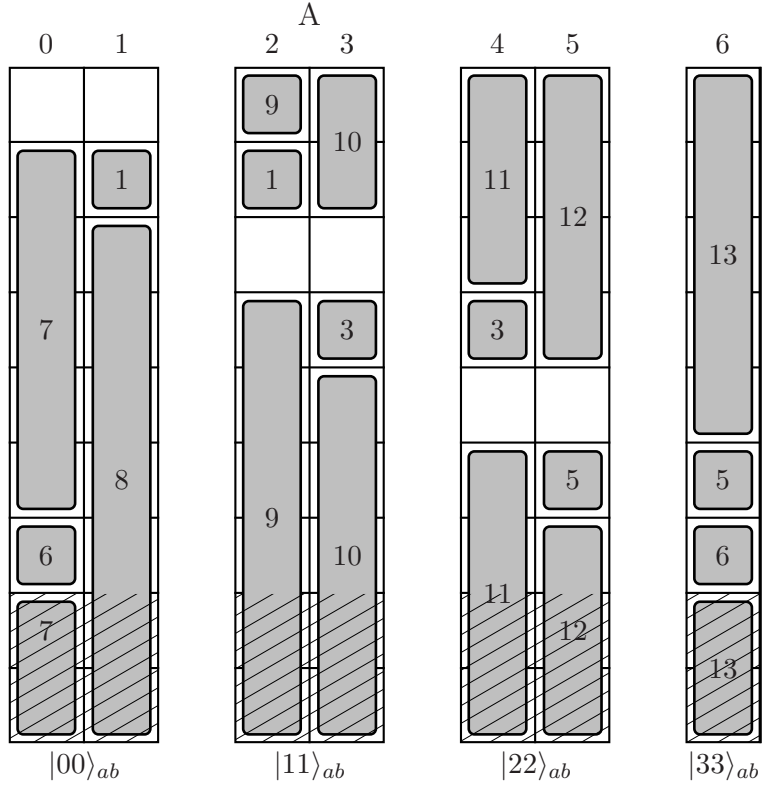}
\caption{\label{fig:GenTiles2_slide7} GenTiles2 for a $7\otimes n$ system after Bob's measurement has removed all but one of the even-numbered short tiles.}
\end{figure}

Things are slightly more complicated when $m$ is odd, since a single even-numbered short tile (that with index $m-1$) remains after case (2) of Bob's first measurement (``preserves everything else"). For example, in the $7\otimes n$ case of Fig.~\ref{fig:GenTiles2_slide7}, tiles $6$ and $8$ have parts that remain in the same row (see columns $0$ and $1$), implying that tile $6$ could not have been separated from the rest by Bob's measurement (since Bob's measurements always preserve whole rows). In the general case of odd $m$, then, even-numbered short tile $m-1$ remains and one of the links of the horizontal chain of tiles (that between the last two short tiles, $m-2$ and $m-1$) is yet to be broken. That is, the last column is still linked to the one before it by tile $m-2$, and to the first column by tile $m-1$. Therefore, Alice's measurement must include one outcome that is a rank-$3$ projector, preserving the first column along with the last two (all other outcomes are as for the even $m$ case, and proceed accordingly). For example, in Fig.~\ref{fig:GenTiles2_slide7}, this outcome would preserve columns $0$, $5$, and $6$ (linked by short tiles $5$ and $6$). Notice, however, that no two of these columns were initially paired with each other in Alice's first ``sliding" measurement. Hence, they have all been slid vertically apart from each other so that they have no horizontal overlap (that is, they are orthogonal on Bob's side), and it is clear that even for this outcome, Bob can separate the two short tiles from each other and from the vertical tiles, after which Alice can either distinguish within the remaining short tile or isolate the vertical tiles from one another. If it is the latter, then Bob can distinguish amongst the states in the remaining vertical tile.

\begin{figure}
\includegraphics{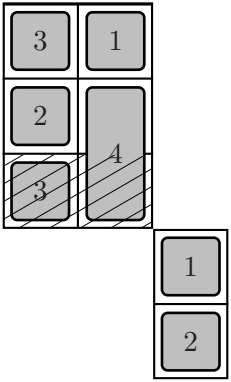}
\caption{\label{fig:GenTiles2_3x3} GenTiles2 on $3\otimes3$ after Alice and Bob have both done initial measurements.}
\end{figure}

This discussion of the odd $m$ case indicates that for $m=3$, there will be a problem. Even though the $0$ tile (and also the one vertical tile in the last -- in this case, third -- column) can be removed by Bob after Alice uses the entanglement to slide the last column down away from the other two, there will remain a link between the other two horizontal tiles, which in this case together span the complete set of Alice's columns. The situation at this stage will be as depicted in Fig.~\ref{fig:GenTiles2_3x3}. We see that the horizontal tiles $1$ and $2$, while broken across the two images, still link each other. Because of this, and since the parties must preserve each tile in its entirety, it would appear as though there is little progress they can make from this point. Thus, apparently, our protocol fails to distinguish the $3\otimes n$ case of GenTiles2, which it appears will require a $3\otimes3$ MES to be distinguished, enough entanglement to allow Alice to teleport her state to Bob.
In all other ($m\ne3$) cases, however, we have shown that with an $\lceil m/2\rceil\otimes\lceil m/2\rceil$ MES, the GenTiles2 UPB can be distinguished by LOCC. \hspace{\stretch{1}}$\blacksquare$


\begin{thebibliography}{10}

\bibitem{BennettTele}
C. Bennett {\it et~al.}, Phys. Rev. Lett. {\bf 70},  1895  (1993).

\bibitem{BennettDense}
C.~H. Bennett and S.~J. Wiesner, Phys. Rev. Lett. {\bf 69},  2881  (1992).

\bibitem{Shor_expand}
P.~W. Shor, SIAM J. Sci. Statist. Comput. {\bf 26},  1484–1509  (1997).

\bibitem{Bennett9}
C.~H. Bennett {\it et~al.}, Phys. Rev. A {\bf 59},  1070  (1999).

\bibitem{GroismanVaidman}
B. Groisman and L. Vaidman, J. Phys. A: Math. Gen. {\bf 34},  6881  (2001).

\bibitem{WalgateHardy}
J. Walgate and L. Hardy, Phys. Rev. Lett. {\bf 89},  147901  (2002).

\bibitem{myLDPE}
S.~M. Cohen, Phys. Rev. A {\bf 75},  052313  (2007).

\bibitem{NisetCerf}
J. Niset and N.~J. Cerf, Phys. Rev. A {\bf 74},  52103  (2006).

\bibitem{IBM_PRL}
C.~H. Bennett {\it et~al.}, Phys. Rev. Lett. {\bf 82},  5385  (1999).

\bibitem{IBM_CMP}
D.~P. DiVincenzo {\it et~al.}, Commun. Math. Phys. {\bf 238},  379–410  (2003).

\bibitem{AlonLovasz}
N. Alon and L. Lovasz, J. Combin. Theory, Series A {\bf 95},  169–179  (2001).

\bibitem{Rinaldis}
S. De~Rinaldis, Phys. Rev. A {\bf 70},  022309  (2004).

\bibitem{HorodeckiBound}
P. Horodecki, Phys. Lett. A {\bf 232},  333  (1997).

\bibitem{HorodeckisBound}
P. Horodecki, M. Horodecki, and R. Horodecki, Phys. Rev. Lett. {\bf 82},  1056
  (1999).

\bibitem{DiVincenzoTerhal}
D. P. DiVincenzo and B. M. Terhal, arXiv:quant-ph/0008055.

\bibitem{NoCloning}
W. Wootters and W. Zurek, Nature {\bf 299},  802  (1982).

\bibitem{cloneDieks}
D. Dieks, Physics Letters A {\bf 92},  271  (1982).

\bibitem{my_tele}
S. M. Cohen, arXiv:0704.0051v1 [physics.ed-ph].

\bibitem{refThnx}
I am grateful to an anonymous referee for raising this interesting question.

\end{thebibliography}

\end{document}